\documentclass[12pt]{article}
\usepackage[pdftex]{graphicx}
 \DeclareGraphicsExtensions{.pdf,.jpeg,.png}

\newcommand{\spr}{*^\prime}
\newcommand{\lb}{{\bf L}}
\newcommand{\bw}{{\bf w}}
\newcommand{\bv}{{\bf v}}
\newcommand{\ze}{{\bf 0}}
\newcommand{\sr}{{\cal R}}
\newcommand{\bu}{{\bf u}}
\newcommand{\beq}{\begin{eqnarray*}}
\newcommand{\eeq}{\end{eqnarray*}}
\newcommand{\llra}{\Longleftrightarrow}
\newcommand{\ck}{{\cal K}}
\newcommand{\gb}{{\bf G}}
\newcommand{\jb}{{\bf J}}
\newcommand{\ib}{{\bf I}}
\newcommand{\hb}{{\bf H}}
\newcommand{\tb}{{\bf T}}
\newcommand{\bx}{{\bf x}}
\newcommand{\by}{{\bf y}}
\newcommand{\pf}{{\bf P}}
\newcommand{\xb}{{\bf X}}
\newcommand{\qb}{{\bf Q}}
\newcommand{\mb}{{\bf M}}
\newcommand{\pLambda}{{\bf \Lambda}}
\newcommand{\diag}{{\rm diag}\,}
\newcommand{\trace}{{\rm trace}\,}
\newcommand{\bet}{\begin{theorem}}
\newcommand{\eet}{\end{theorem}}
\newcommand{\rank}{{\rm rank}\,}

\newcommand{\xbp}{{\xb^\prime}}
\newcommand{\E}{{\rm E}\,}
\newcommand{\Var}{{\rm Var}\,}
\newcommand{\pPhi}{{\bf \Phi}}
\newcommand{\yb}{{\bf Y}}
\newcommand{\byp}{{\by^\prime}}
\newcommand {\bfb}{{\mbox{\boldmath $\beta$}}}
\newcommand {\bft}{{\mbox{\boldmath $\tau$}}}
\newcommand {\bfe}{{\mbox{\boldmath $\eta$}}}
\newcommand{\zb}{{\bf Z}}
\newcommand{\zbp}{{\zb^\prime}}
\newcommand {\nnp}{{\bf NN^\prime}}
\newcommand{\bys}{{\bf y^*}}
\newcommand{\bysp}{{\bys^\prime}}

\begin{document}

\begin{titlepage}
    \begin{center}
        \vspace*{1cm}
            
        \Huge
        \textbf{Projection matrices and the sweep operator}

        \vspace{2.5cm}
         \huge   
        \textbf{
A.T. James$^a$ and
E.R.Williams$^b$ \footnote{ Author to whom correspondence should be
addressed\\ \indent $^a$ The University of Adelaide, SA 5005 Australia  \\
\indent $^b$ The Australian National University, ACT 2600, Australia  \\ \indent e-mail:
emlyn@alphastat.net \\ }}

\vspace {1 cm}
\large {\em The University of Adelaide and The Australian National University}

    \end{center}
\end{titlepage}

\begin{center}
\large{\bf \ \ Foreword}\\
\end{center}
\noindent{These notes have been adapted from an undergraduate course given by Professor Alan James at the University of Adelaide from around 1965 and onwards.  This adaption has put a focus on the definition of projection matrices and the sweep operator.  These devices were at the heart of the development of the statistical package Genstat.  The first version focussed on the analysis of variance using the sweep operator.  Later on, after spending a period with Graham Wilkinson at CSIRO in Adelaide, John Nelder expanded the package to include generalized linear models.
I was fortunate to have Professor James as my Honours supervisor before completing my PhD under the supervision of Professor H.D. Patterson.  It is ironic that Professor James (the birth of Genstat), Professor Patterson (the birth of REML)  and my Father (the birth of me!) were all born within five days of each other in July 1924.  Hence, if still alive, they would have celebrated their 100th birthday this year.
I am thankful to Professor James for the algebra behind projection matrices and the use of the sweep operator which has since been applied in a range of situations, e.g. see [1…5] below.}
  
\noindent{Emlyn Williams}

\noindent{April 2024}
\vspace {1 cm}

\noindent{[1] Williams, E.R. (1986). A neighbour model for field experiments. {\em Biometrika 73}, 279-87.}

\noindent{[2] De Hoog, F.R., Speed, T.P. and Williams, E.R. (1990). On a matrix identity associated with generalized least squares. {\em Lin. Alg. Appl. 127}, 449-56.}

\noindent{[3] Piepho, H.P., Williams, E.R. and Madden, L.V. (2012). The use of two-way linear mixed models in multi-treatment meta-analysis. {\em Biometrics 68}, 1269-77.}

\noindent{[4] Boer, M.P., Piepho, H.P. and Williams, E.R. (2020). Linear Variance, P-splines and neighbour differences for spatial adjustment in field trials: How are they related? {\em J. Agric., Biol. and Environ. Statist. 25}, 676-98.}

\noindent{[5] Piepho, H.P., Williams, E.R., Harwood, C.E. and Prus, M. (2024). Assessing the efficiency and heritability of blocked tree breeding trials. (in progress).}

\newpage

\begin{center}
\large{\bf \ \ Professor Alan Treleven James (1924-2013)}
                      
\normalsize{\bf  (Adapted from Alan James' eulogy, 2013 by the University of Adelaide)}\\
\end{center}
\vspace{0.25cm}

Born in Berri, South Australia, in 1924, Alan Treleven James was the youngest child of litigant and dried fruit merchant, Frederick Alexander James and Rachel May James. Alan James attended the Glen Osmond Primary School and later won the Samuel Fiddian scholarship to attend Prince Alfred College. In 1944 he completed a Bachelor of Science with Honours at the University of Adelaide and a Masters of Science in 1949.

Alan James' first professional job with CSIR, now CSIRO, included teaching throughout his early twenties at The University of Adelaide. During a posting to CSIR in Canberra he met a colleague, Cynthia, who would become is wife in 1950. Shortly afterwards, Alan James was awarded a CSIRO studentship to study at Princeton University, New Jersey, where he completed his PhD in 1952. Alan James and Cynthia returned to Adelaide and their first two children, Michael and Stephen, were born.

Alan James continued to work for CSIRO until 1958, when he resigned to take up a one year teaching position at Yale University in Connecticut. He progressed to full professorship and he and Cynthia had two more sons, Andrew and Nicholas.

In 1965, Alan James and his family returned to Adelaide again, where he took on the role of Chair of Mathematical Statistics at The University of Adelaide. Although he retired from the University in 1989, Alan James continued his research activities for many years. In fact, he was awarded the 1992 Pitman Medal of the Statistical Society of Australia in recognition of his outstanding achievement and contribution to statistics, particularly in the field of multivariate analysis.

Alan James also loved to travel through the Australian outback and took his family on many adventurous car trips to Maree and beyond. On one trip, the family's station wagon would become stuck in floods south of Innamincka for a week. He also loved to make wine and to dry fruit, following in the footsteps of his father's profession.
Alan James was an active church-goer for most of his life and a member of the St John's Anglican Church since 1980. He served on the Parish Council and the boards of the St John's Youth Services and the Magdelene Centre.
In 2002, Alan James was diagnosed with cancer. Although ongoing treatments won him several years of remission, he died of a heart attack on 7th March 2013, after a period of deteriorating health.  He is survived by his wife, four children and nine grandchildren.
Alan James' publications include approximately forty research papers, those published in the Annals of Mathematics between 1954 and 1961, laying the foundation for much of his later work in multivariate analysis, as well as for the work of many others.
\newpage

\begin{center}
\large{\bf \ \ 1. Linear Models}
\end{center}
\vspace{0.25cm}

\noindent
{\bf 1.1 The model}\\
The purpose of a linear model is to summarize a vector of observations 
$\by\in\sr^n$ with an approximating vector $\yb=\xb\mbox{\boldmath $\pi$}$, where $\mbox{\boldmath $\pi$}\in\sr^p$ 
is a vector of $p$ parameters which have to be estimated and $\xb$ is an 
$n\times p$ matrix of constants. The model restricts the sample space $\sr^n$ 
to the subspace $\sr(\xb)$ which is known as the model subspace. The difference 
$\by-\yb$ between the observed and approximating vectors is called the residual 
or error vector $\mbox{\boldmath $\epsilon$}$, thus the linear model becomes 
\begin{equation}
\by=\xb\mbox{\boldmath $\pi$}+\mbox{\boldmath $\epsilon$}\
\label {eq:1.1}
\end{equation}

In multiple regression analysis the columns of $\xb$ represent $p$ independent 
variates and are usually linearly independent, i.e.\ rank$\,(\xb)=p$. However 
for designed experiments the columns of $\xb$ are dummy  variates consisting of 
zeros and ones. It is then more natural to retain as much symmetry as possible 
in the model, usually at the expense of introducing linear dependence between 
the columns of $\xb$. Then rank$\,(\xb)$ is less than $p$ and the value $\mbox{\boldmath $\pi$}$ 
of the parameter vector which yields a certain approximating vector $\yb$ is 
not unique. In fact all vectors in the coset $\mbox{\boldmath $\pi$}+\ck(\xb)$ give the same 
vector $\yb$; the quotient space $\sr^p/\ck(\xb)$ is known as the parameter 
space of the linear model. The only meaningful linear functions of the elements 
of $\mbox{\boldmath $\pi$}$ are those which are unique for each coset in the parameter space.

\noindent{\it Definition 1.1.} A linear function $\mbox{\boldmath $\eta$}^\prime\mbox{\boldmath $\pi$}$ of the 
elements of $\mbox{\boldmath $\pi$}$ is called an estimable linear function if $\mbox{\boldmath $\eta$}$ is 
orthogonal to $\ck(\xb)$.

An estimable linear function $\mbox{\boldmath $\eta$}^\prime\mbox{\boldmath $\pi$}$ is unique for any vector in the 
coset $\mbox{\boldmath $\pi$}+\ck(\xb)$. When rank$\,(\xb)=p$, \ $\ck(\xb)=\ze_p$ and the 
parameter space becomes $\sr^p$; thus all linear functions $\mbox{\boldmath $\eta$}^\prime\mbox{\boldmath $\pi$}$ 
are estimable. 

\bigskip
\noindent
{\bf 1.2 Least squares estimation}\\
Given $\xb$, a suitable approximating vector $\yb$ can be obtained by using 
$\by$ to estimate the elements of $\mbox{\boldmath $\pi$}$. The best known criterion is the 
method of least squares where the parameters $\pi_i$ ($i=1,\ldots,p$) are 
estimated by the quantities that minimize the scalar product
\begin{equation}
\mbox{\boldmath $\epsilon$}^\prime\mbox{\boldmath $\epsilon$}=(\by-\xb\mbox{\boldmath $\pi$})^\prime(\by-\xb\mbox{\boldmath $\pi$}).\
\label {eq:1.2}
\end{equation}

\noindent{\it Theorem 1.1.} The vector $\hat{\mbox{\boldmath $\pi$}}$ of least squares estimators of the 
elements of $\mbox{\boldmath $\pi$}$ is a solution of the equations 
\begin{equation}
(\xb^\prime\xb)\hat{\mbox{\boldmath $\pi$}}=\xb^\prime\by.\
\label {eq:1.3}
\end{equation}

\noindent{\it Proof.} The vector $\hat{\mbox{\boldmath $\pi$}}$ which minimizes (\ref{eq:1.2}) is obtained 
by differentiation with respect to $\mbox{\boldmath $\pi$}$ and equating the result to zero. Thus
\[\frac{d\mbox{\boldmath $\epsilon$}^\prime\mbox{\boldmath $\epsilon$}}{d\mbox{\boldmath $\pi$}}=-2\xb^\prime(\by-\xb\hat{\mbox{\boldmath $\pi$}})=\ze_p\;,\]
and hence
\[(\xb^\prime\xb)\hat{\mbox{\boldmath $\pi$}}=\xb^\prime\by\;.\]
The equations (\ref{eq:1.3}) are called the normal equations. Their solution depends on 
the relationship of $\xb$ to $\xb^\prime\xb$.

\noindent{\it Theorem 1.2.} If $\xb$ is an $n\times p$ matrix then 
\begin{enumerate}
\item[(i)] $\ck(\xb^\prime\xb)=\ck(\xb)$
\item[(ii)] ${\rm rank}\,(\xb^\prime\xb)={\rm rank}\,(\xb)$
\item[(iii)] $\sr(\xb^\prime\xb)=\sr(\xb^\prime)\;.$
\end{enumerate}

\noindent{\it Proof.} (i) Suppose $\mbox{\boldmath $\pi$}\in\ck(\xb^\prime\xb)$, then 
\begin{eqnarray*} &&\xb^\prime\xb\mbox{\boldmath $\pi$}=\ze_p\\
&\llra&\mbox{\boldmath $\pi$}^\prime\xb^\prime\xb\mbox{\boldmath $\pi$}=0\\
&\llra&(\xb\mbox{\boldmath $\pi$})^\prime(\xb\mbox{\boldmath $\pi$})=0\\
&\llra&\xb\pi=\ze_n\qquad\qquad\mbox{(Theorem A.16)}\\
&\llra&\xb^\prime\xb\mbox{\boldmath $\pi$}=\ze_p\;.
\end{eqnarray*}
Hence $\mbox{\boldmath $\pi$}\in\ck(\xb^\prime\xb)$ if and only if $\mbox{\boldmath $\pi$}\in\ck(\xb)$, 
\[{\rm i.e.\ }\qquad\ck(\xb^\prime\xb)=\ck(\xb)\;.\]

(ii) It follows from Theorems A.13 and A.14 that 
\begin{eqnarray*} \rank(\xb^\prime\xb)
&=&p-\dim\bigl(\ck(\xb^\prime\xb)\bigr)\\
&=&p-\dim\bigl(\ck(\xb)\bigr)\qquad\mbox{(from (i))}\\
&=&\rank(\xb)\;.\end{eqnarray*}

(iii) Clearly $\sr(\xb^\prime\xb)\subset\sr(\xb^\prime)$, but 
$\dim(R(\xb^\prime\xb))=\rank(\xb^\prime\xb)=\rank(\xb)$, and from Theorem A.11,
\begin{eqnarray*} \rank(\xb)
&=&\rank(\xb^\prime)\\ 
&=&\dim\bigl(\sr(\xb^\prime)\bigr)\;.
\end{eqnarray*}
Hence $\dim\bigl(\sr(\xb^\prime\xb)\bigr)=\dim\bigl(\sr(\xb^\prime)\bigr)$ and 
so 
\[\sr(\xb^\prime\xb)=\sr(\xb^\prime)\;.\]

Hence when the columns of $\xb$ are linearly independent, $\xb^\prime\xb$ is 
non-singular and the solution of the normal equations is unique and given by 
\[\hat{\mbox{\boldmath $\pi$}}=(\xb^\prime\xb)^{-1}\xb^\prime\by\;.\]
However if the rank of $\xb$ is less than $p$, \ $\xb^\prime\xb$ will be singular
 and a generalized inverse for $\xb^\prime\xb$ is needed to solve the normal 
equations. For example suppose $\gb$ is a generalized inverse of 
$\xb^\prime\xb$, then a solution of (\ref{eq:1.3}) is 
\begin{equation}\hat{\mbox{\boldmath $\pi$}}=\gb\xb^\prime\by\;.
\end{equation}
This solution is not unique but dependent on the choice of $\gb$. The full 
solution to the normal equations is the coset $\hat{\mbox{\boldmath $\pi$}}+\ck(\xb)$ in the 
parameter space $\sr^p/\ck(\xb)$. The least squares approximating vector 
$\yb=\xb\hat{\mbox{\boldmath $\pi$}}$, and least squares estimator $\mbox{\boldmath $\eta$}^\prime\hat{\mbox{\boldmath $\pi$}}$ of an 
estimable linear function $\mbox{\boldmath $\eta$}^\prime\mbox{\boldmath $\pi$}$ are unique for any vector in 
$\hat{\mbox{\boldmath $\pi$}}+\ck(\xb)$. 

The analysis of variance for the model (\ref{eq:1.1}) is given in Table \ref{table:1.1}, where 
$q=\rank(\xb)$. 

\begin{table}
\centering\small
\caption{{\it Analysis of Variance}}
\vspace{0.2cm}
\begin{tabular}{lcc}
\hline
&{\bf df}&{\bf ss}\\
\multicolumn{1}{c}{\it Units stratum}&&\\
Model&$q$&$\yb^\prime\yb=\hat{\mbox{\boldmath $\pi$}}^\prime\xbp\by$\\
Residual&$n-q$&by difference\\
\hline
Grand Total&$n$&$\by^\prime\by$\\
\hline
\end{tabular}
\label{table:1.1}
\end{table}

\bigskip
\noindent{\bf 1.3 Projections on the model subspace}\\
For the linear model (\ref{eq:1.1}) a projection of $\sr^n$ on $\sr(\xb)$ will map $\by$ 
to an approximating vector $\yb=\xb\mbox{\boldmath $\pi$}$. Different projections on $\sr(\xb)$ 
will give different $\yb$ vectors. The following theory establishes that the 
approximating vector $\yb=\xb\hat{\mbox{\boldmath $\pi$}}$, where $\hat{\mbox{\boldmath $\pi$}}$ is a least squares 
estimator of $\mbox{\boldmath $\pi$}$, is obtained from the orthogonal projection of $\sr^n$ on 
$\sr(\xb)$. 

\noindent{\it Theorem 1.3.} If $\xb$ is an $n\times p$ matrix then a square matrix 
$\gb$ of order $p$ is a generalized inverse of $\xb^\prime\xb$ if and only if 
$\gb\xb^\prime$ is a generalized inverse of $\xb$, i.e.
\[\xb^\prime\xb\gb\xb^\prime\xb=\xb^\prime\xb\llra\xb\gb\xb^\prime\xb=\xb\;.\]

\noindent{\it Proof.} (i) If $\xb\gb\xb^\prime\xb=\xb$ then clearly
\[\xb^\prime\xb\gb\xb^\prime\xb=\xb^\prime\xb.\]

(ii) Suppose now that $\xb^\prime\xb\gb\xb^\prime\xb=\xb^\prime\xb$. From 
Theorems A.18 and A.20 it follows that 
\[\sr^n=\sr(\xb)\oplus\ck(\xb^\prime)\;.\]
Hence for any $\by\in\sr^n$, there exist vectors $\mbox{\boldmath $\pi$}\in\sr^p$ and 
$\by_0\in\ck(\xb^\prime)$ such that 
\[\by=\xb\mbox{\boldmath $\pi$}+\by_0\;.\]
Therefore 
\begin{eqnarray*}
\by^\prime\xb\gb\xb^\prime\xb
&=&\mbox{\boldmath $\pi$}^\prime\xb^\prime\xb\gb\xb^\prime\xb+\by_0^\prime\xb\gb\xb^\prime\xb\\
&=&\mbox{\boldmath $\pi$}^\prime\xb^\prime\xb+\ze^\prime_p\\
&=&\by^\prime\xb
\end{eqnarray*}
for all $\by\in\sr^n$. Thus \[\xb\gb\xb^\prime\xb=\xb\;.\]

\noindent{\it Theorem 1.4.} If $\gb$ is a generalized inverse of $\xb^\prime\xb$ then
\begin{eqnarray*}
&{\rm (i)}\quad&\xb^\prime\xb\gb\xb^\prime=\xb^\prime.\\
&{\rm (ii)}\quad&\xb\gb^\prime\xb=\xb\gb\xb^\prime\;,\quad\mbox{i.e.\ 
$\xb\gb\xb^\prime$  is symmetric}\;.
\end{eqnarray*}

\noindent{\it Proof.} (i) This follows by calculating 
$\xb^\prime\xb\gb\xb^\prime\by$ in a manner similar to Theorem 1.3.

(ii) Taking the transpose of (i) gives 
\[\xb\gb^\prime\xb^\prime\xb=\xb\;,\]
and using Theorem 1.3, 
\[\xb\gb^\prime\xb^\prime\xb=\xb\gb\xb^\prime\xb\;.\]
Hence 
\[\xb\gb^\prime\xb^\prime(\xb\mbox{\boldmath $\pi$}+\by_0)=\xb\gb\xb^\prime(\xb\mbox{\boldmath $\pi$}+\by_0)\]
for some $\mbox{\boldmath $\pi$}\in\sr^p$ and $\by_0\in\ck(\xb^\prime)$. But since any 
$\by\in\sr^n$ can be written as $\xb\mbox{\boldmath $\pi$}+\by_0$, 
\[\xb\gb^\prime\xb\by=\xb\gb\xb^\prime\by\qquad\mbox{for all 
$\by\in\sr^n$}\;,\]
\[{\rm i.e.}\qquad\xb\gb^\prime\xb^\prime=\xb\gb\xb^\prime\;.\]

\noindent{\it Theorem 1.5.} If $\gb$ is a generalized inverse of $\xbp\xb$ then 
$\xb\gb\xbp$ is \ (i) idempotent and \ (ii) has range $\sr(\xb)$.

\noindent{\it Proof.} \begin{eqnarray*}
{\rm (i)}\qquad(\xb\gb\xbp)^2
&=&(\xb\gb\xbp\xb)\gb\xbp\\
&=&\xb\gb\xbp,\qquad\mbox{(from Theorem 1.3)}
\end{eqnarray*}
i.e. $\xb\gb\xbp$ is idempotent.

(ii) Clearly $\sr(\xb\gb\xbp)\subset\sr(\xb)$.
Now suppose $\yb\in\sr(\xb)$, i.e.\ $\yb=\xb\mbox{\boldmath $\pi$}$ for some $\mbox{\boldmath $\pi$} \in\sr^p$. 
Then from Theorem 1.3, \[\yb=\xb\gb\xbp\xb\mbox{\boldmath $\pi$}=\xb\gb\xbp\yb\;.\]
Thus $\yb\in\sr(\xb\gb\xbp)$ and hence 
\[\sr(\xb\gb\xbp)=\sr(\xb)\;.\]

\noindent{\it Theorem 1.6.} If $\gb$ is a generalized inverse of $\xbp\xb$ then 
$\xb\gb\xbp$ is the matrix of the orthogonal projection on $\sr(\xb)$. 

Since $\pf=\xb\gb\xbp$ is the matrix of an orthogonal projection on $\sr(\pf)$, 
it follows from Theorem A.5 that $\pf$ is unique. Thus if $\gb^*$ is any other 
generalized inverse of $\xbp\xb$ then 
\[\xb\gb\xbp=\xb\gb^*\xbp\;.\]

\noindent{\it Theorem 1.7.} Let $\pf=\xb\gb\xbp$ be the matrix of the orthogonal 
projection of $\sr^n$ on $\sr(\xb)$ then $\xb\hat{\mbox{\boldmath $\pi$}}=\pf\by$ if and only if 
$\hat{\mbox{\boldmath $\pi$}}$ is a least squares estimator of $\mbox{\boldmath $\pi$}$. 

\noindent{\it Proof.} (i) Let $\xb\hat{\mbox{\boldmath $\pi$}}=\pf\by$ then 
\begin{eqnarray*} \xbp\xb\hat{\mbox{\boldmath $\pi$}}
&=&\xbp\pf\by\\
&=&\xbp\by,\qquad\mbox{(by Theorem 1.4)}
\end{eqnarray*}
i.e.\ $\hat{\mbox{\boldmath $\pi$}}$ satisfies the normal equations and is thus a least squares 
estimator. 

(ii) Multiplying $\xbp\xb\hat{\mbox{\boldmath $\pi$}}=\xbp\by$ by $\xb\gb$ gives 
\[\xb\gb\xbp\xb\hat{\mbox{\boldmath $\pi$}}=\xb\gb\xbp\by\;,\]
i.e. \ \ \ \ \ \ \ \ \ $\xb\hat{\mbox{\boldmath $\pi$}}=\pf\by$.

Figure 1 gives a geometrical interpretation of the orthogonal projection on 
$\sr(\xb)$. 

\begin{figure}
	\centering
		\includegraphics [scale=0.5]{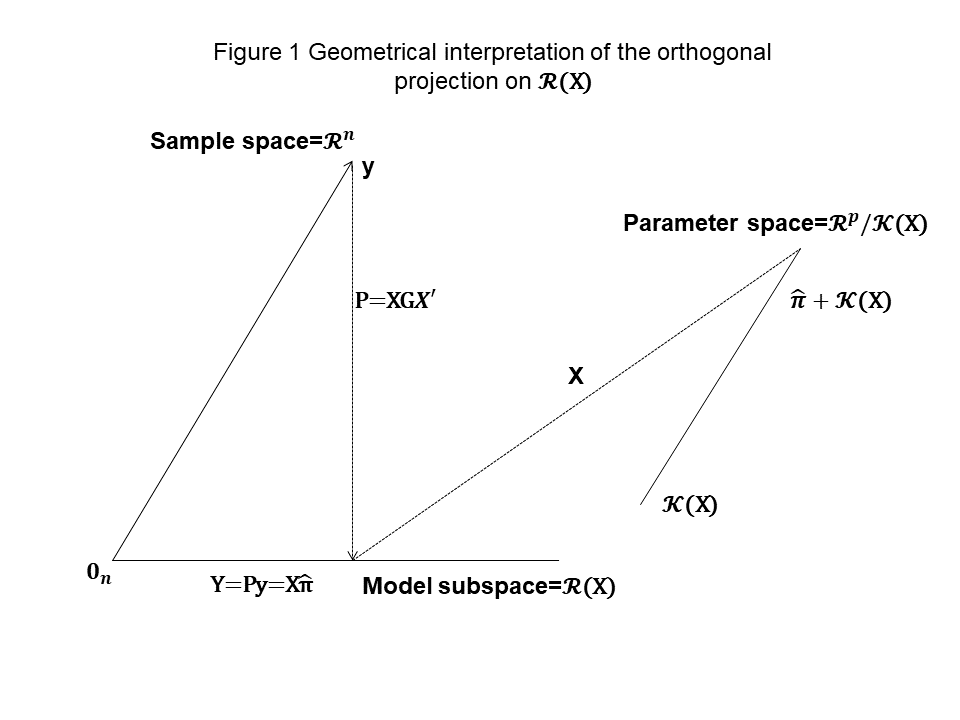}
	\label{fig:1}
\end{figure}

In the analysis of variance (Table \ref{table:1.1}), the model sum of squares 
$\yb^\prime\yb$ can be rewritten as 
\begin{eqnarray*}
\yb^\prime\yb
&=&\hat{\mbox{\boldmath $\pi$}}^\prime\xbp\by\\
&=&\by^\prime\xb\gb\xbp\by\\
&=&\by^\prime\pf\by\;.
\end{eqnarray*}
Table \ref{table:1.2} then gives an alternate form for the analysis of variance. 

\begin{table}
\centering\small
\caption{{\it Analysis of Variance}}
\vspace{0.2cm}\begin{tabular}{lcc}
\hline
&{\bf df}&{\bf ss}\\
\multicolumn{1}{c}{\it Units stratum}&&\\
Model&$q$&$\by^\prime\pf\by$\\
Residual&$n-q$&$\by^\prime(\ib_n-\pf)\by$\\
\hline
Grand Total&$n$&$\by^\prime\by$\\
\hline
\end{tabular}
\label{table:1.2}
\end{table}

\bigskip
\noindent{\bf 1.4 Distributional assumptions}\\
So far the discussion of the linear model (\ref{eq:1.1}) has dealt with the 
approximation of $\by$ by a vector $\yb\in\sr(\xb)$. In order to decide which 
model best represents the data, it is necessary to make certain statistical 
assumptions. These will be progressively introduced in this section. Initially 
the observations $y_1,\ldots,y_n$ are taken to be random variates with 
probability density function $f(y_1,\ldots,y_n)=f(\by)$. 

\noindent{\it Definition 1.2.} If $y_1,\ldots,y_n$ are jointly distributed 
random variates with a probability density function $f(\by)$, and if 
$\phi(\by)$ is a function of $y_1,\ldots,y_n$ then the expectation of $\phi$ is 
\[\E[\phi]=\int_{y_1}\ldots\int_{y_n}\phi(\by)f(\by)\,dy_1,\ldots,dy_n\;.\]

\noindent{\it Definition 1.3.} Let $\pPhi$ be an $l\times m$ matrix variate 
with elements $\Phi_{ij}(\by)$, ($i=1,\ldots,l$;  $j=1,\ldots,m$).
The expectation of $\pPhi$ is defined as the matrix of expectations
\[\E[\pPhi]=\bigl\{\E[\Phi_{ij}]\bigr\}\;.\]
The following properties of expectations can then be stated.

\noindent{\it Theorem 1.8.} If $\pPhi_1$ and $\pPhi_2$ are $l\times m$ matrix variates 
then 
\[\E[\pPhi_1+\pPhi_2]=\E[\pPhi_1]+\E[\pPhi_2]\;.\]

\noindent{\it Theorem 1.9.}
If $\pPhi$ is an $l\times m$ matrix variate and $\lb$ and $\mb$ are 
$l^\prime\times l$ and $m\times m^\prime$ matrices of constants respectively 
then 
\[\E[\lb\pPhi\mb]=\lb\E[\pPhi]\mb\;.\]

\noindent{\it Definition 1.4.} The variance matrix $\Var(\by)$ of a vector 
variate $\by\in\sr^n$ is a square matrix of order $n$ given by
\[\Var(\by)=\E\left[(\by-E[\by])(\by-\E[\by])^\prime\right]\;.\]
The $(i,i)$th element of $\Var(\by)$ is the variance of the $i$th element of 
$\by$. The $(i,j)$th element of $\Var(\by)$, \ $i\ne j$ is called the 
covariance between the $i$th and $j$th elements of $\by$.

\noindent{\it Theorem 1.10.} If $\by\in\sr^n$ is a vector variate and $\lb$ is a 
$p\times n$ matrix of constants then 
\[\Var(\lb\by)=\lb\Var(\by)\lb^\prime\;.\]

\noindent{\it Proof.}
From Definition 1.4 
\begin{eqnarray*}\Var(\lb\by)
&=&\E\left[(\lb\by-\E[\lb\by])(\lb\by-\E[\lb\by])^\prime\right]\\
&=&\E\left[\lb(\by-\E[\by])(\by-\E[\by])^\prime\lb^\prime\right]\quad\mbox{
(by Theorem 1.9)}\\
&=&\lb\E\left[(\by-\E[\by])(\by-\E[\by])^\prime\right]\lb^\prime\\
&=&\lb\Var(\by)\lb^\prime\;.
\end{eqnarray*}
Assumptions about the expectation and variance matrix of the vector variate 
$\by$ in the linear model (\ref{eq:1.1}) are now added, namely
\begin {equation}
\E[\by]=\xb\mbox{\boldmath $\pi$}\;;\qquad\Var(\by)=\sigma^2\ib_n\;,\hfill\
\label{eq:1.5}
\end {equation}
where $\sigma^2$ is a further parameter which usually has to be estimated from 
the data. An equivalent form for (\ref{eq:1.5}) is to write the assumptions as 
\begin {equation}
\E[\mbox{\boldmath $\epsilon$}]=\ze_n\;;\quad\Var(\mbox{\boldmath $\epsilon$})=\E[\mbox{\boldmath $\epsilon$}\mbox{\boldmath $\epsilon$}^\prime]=\sigma^2\ib_n\;,
\hfill\
\label{eq:1.6}
\end {equation}
i.e.\ the elements of the error vector each have zero mean, variance equal to 
$\sigma^2$ and are uncorrelated.

The assumptions (\ref{eq:1.5}) on the linear model (\ref{eq:1.1}) partition $\by$ into two 
components, \ (i) a fixed component $\E[\by]=\yb$ in the model subspace 
$\sr(\xb)$ and \ (ii) a random or error component distributed with zero 
expectation and variance matrix $\sigma^2\ib_n$. 

Later on a further assumption will be needed for the purpose 
of testing hypotheses about the linear model. Then the probability density 
function $f(\by)$ will be assumed to be multivariate normal, i.e.
\begin {equation}
\by\sim N(\xb\mbox{\boldmath $\pi$}\,,\,\sigma^2\ib_n)\;,\
\label{eq:1.7}
\end {equation}
or equivalently
\[\mbox{\boldmath $\epsilon$}\sim N(\ze_n\,,\,\sigma^2\ib_n)\;.\hfill\]

\bigskip
\noindent{\bf 1.5 Properties of least squares estimators}\\
The least squares estimator $\mbox{\boldmath $\eta$}^\prime\gb\xb^\prime\by$ of an estimable 
linear function $\mbox{\boldmath $\eta$}^\prime\mbox{\boldmath $\pi$}$ is just one of a class of linear estimators 
of the form \[\mbox{\boldmath $\eta$}^\prime\lb\by\;,\]
where $\lb$ is a $p\times n$ matrix. However least squares estimators have a 
number of desirable properties, some of which will be demonstrated in this 
section. Firstly some preliminary results are obtained.

\noindent{\it Theorem 1.11.} A $p\times n$ matrix $\lb$ is a generalized inverse of 
$\xb$ if and only if $\xb\lb$ is the matrix of a projection on $\sr(\xb)$.

\noindent{\it Proof.} (i) Suppose $\lb$ is a generalized inverse of $\xb$, i.e.
\[\xb\lb\xb=\xb\;.\]
Then 
\[(\xb\lb)^2=(\xb\lb\xb)\lb=\xb\lb\;,\]
hence $\xb\lb$ is idempotent and using Theorem A.6, $\xb\lb$ is a projection 
matrix with range $\sr(\xb\lb)$. It remains to show that 
$\sr(\xb\lb)=\sr(\xb)$. Clearly $\sr(\xb\lb)\subset\sr(\xb)$. Now suppose 
$\yb\in\sr(\xb)$, i.e.\ $\yb=\xb\mbox{\boldmath $\pi$}$ for some $\mbox{\boldmath $\pi$}\in\sr^p$. Then 
\begin{eqnarray*} \yb
&=&\xb\lb\xb\mbox{\boldmath $\pi$}\qquad\mbox{(by assumption)}\\
&=&\xb\lb\yb\;.
\end{eqnarray*}
Thus $\yb\in\sr(\xb\lb)$ and hence \[\sr(\xb\lb)=\sr(\xb)\].

(ii) Suppose $\xb\lb$ is the matrix of a projection on $\sr(\xb)$, i.e.
\[\xb\lb\yb=\yb\;,\]
for all $\yb\in\sr(\xb)$. Thus 
\[\xb\lb\xb\mbox{\boldmath $\pi$}=\xb\mbox{\boldmath $\pi$}\;,\]
for all $\mbox{\boldmath $\pi$}\in\sr^p$, i.e. 
\[\xb\lb\xb=\xb\;.\]

\noindent{\it Theorem 1.12.} A linear function $\mbox{\boldmath $\eta$}^\prime\mbox{\boldmath $\pi$}$ is estimable if and 
only if $\mbox{\boldmath $\eta$}\in\sr(\xb^\prime)$. 

\noindent{\it Proof.} From Definition 1.1, $\mbox{\boldmath $\eta$}^\prime\mbox{\boldmath $\pi$}$ is estimable 
$\llra\mbox{\boldmath $\eta$}$ is orthogonal to $\ck(\xb)$.

Now 
\beq \ck(\xb)
&=&\ck(\xbp\xb)\qquad\mbox{(by Theorem 1.2)}\\
&=&\sr(\xbp\xb)^\perp\;.\qquad\mbox{(by Theorem A.19)}
\eeq
Hence $\mbox{\boldmath $\eta$}^\prime\mbox{\boldmath $\pi$}$ is estimable
\beq \llra\mbox{\boldmath $\eta$}
&\in&\sr(\xbp\xb)\\
&=&\sr(\xbp)\;.\qquad\mbox{(by Theorem 1.2)}
\eeq

Therefore if $\mbox{\boldmath $\eta$}^\prime\mbox{\boldmath $\pi$}$ is an estimable linear function, then $\mbox{\boldmath $\eta$}$ 
can be written as 
\begin {equation}
\mbox{\boldmath $\eta$}=\xb^\prime\mbox{\boldmath $\gamma$}\;,\hfill\
\label{eq:1.8}
\end{equation}
for some $\mbox{\boldmath $\gamma$}\in\sr^n$. 

\noindent{\it Definition 1.5.} An estimator $a$ of a parametric function 
$\alpha$ is unbiased if 
\[\E[a]=\alpha\;.\]

\noindent{\it Theorem 1.13.} A linear estimator $\mbox{\boldmath $\eta$}^\prime\lb\by$ is an unbiased 
estimator of $\mbox{\boldmath $\eta$}^\prime\mbox{\boldmath $\pi$}$ for all $\mbox{\boldmath $\eta$}$ for which $\mbox{\boldmath $\eta$}^\prime\mbox{\boldmath $\pi$}$ is 
estimable and all $\mbox{\boldmath $\pi$}$ if and only if $\xb\lb$ is the matrix of a projection 
on $\sr(\xb)$. 

\noindent{\it Proof.} $\mbox{\boldmath $\eta$}^\prime\lb\by$ is an unbiased estimator of 
$\mbox{\boldmath $\eta$}^\prime\mbox{\boldmath $\pi$}$
\beq &\llra&\E[\mbox{\boldmath $\eta$}^\prime\lb\by]=\mbox{\boldmath $\eta$}^\prime\mbox{\boldmath $\pi$}\\
&\llra&\mbox{\boldmath $\eta$}^\prime\lb\xb\mbox{\boldmath $\pi$}=\mbox{\boldmath $\eta$}^\prime\mbox{\boldmath $\pi$}\\
&\llra&\mbox{\boldmath $\gamma$}^\prime\xb\lb\xb\mbox{\boldmath $\pi$}=\mbox{\boldmath $\gamma$}^\prime\xb\mbox{\boldmath $\pi$}\quad\mbox{for all 
$\mbox{\boldmath $\gamma$}$ and $\mbox{\boldmath $\pi$}$}\\
&\llra&\xb\lb\xb=\xb\\
&\llra&\xb\lb\quad\mbox{is the matrix of a projection on $\sr(\xb)$ (by Theorem 
1.11)}\;.
\eeq

The least squares estimator $\mbox{\boldmath $\eta$}^\prime\gb\xb^\prime\by$ corresponds to 
setting $\lb=\gb\xb^\prime$, whence the projection matrix $\xb\lb$ in Theorem 
1.13 becomes the orthogonal projection matrix $\pf=\xb\gb\xbp$. The following 
theorem can then be stated:

\noindent{\it Theorem 1.14.} The least squares estimator $\mbox{\boldmath $\eta$}^\prime\gb\xbp\by$ of 
an estimable linear function $\mbox{\boldmath $\eta$}^\prime\mbox{\boldmath $\pi$}$ is unbiased.

Another important property of least squares estimators is given by the 
following result, known as the Gauss-Markov Theorem:

\noindent{\it Theorem 1.15.} The minimum variance unbiased linear estimator of an 
estimable linear function $\mbox{\boldmath $\eta$}^\prime\mbox{\boldmath $\pi$}$ is the least squares estimator 
$\mbox{\boldmath $\eta$}^\prime\gb\xbp\by$, which has variance $\sigma^2\mbox{\boldmath $\eta$}^\prime\gb\mbox{\boldmath $\eta$}$. 

\noindent{\it Proof.} Let $\mbox{\boldmath $\eta$}^\prime\lb\by$ be an unbiased linear estimator 
of an estimable linear function $\mbox{\boldmath $\eta$}^\prime\mbox{\boldmath $\pi$}$. Then from Theorem 1.13, 
$\xb\lb$ is the matrix of a projection on $\sr(\xb)$, but not necessarily the 
orthogonal projection matrix $\pf=\xb\gb\xbp$. The variance of 
$\mbox{\boldmath $\eta$}^\prime\lb\by$ is given by
\beq\Var(\mbox{\boldmath $\eta$}^\prime\lb\by)
&=&\Var(\mbox{\boldmath $\gamma$}^\prime\xb\lb\by)\qquad\mbox{(from (\ref{eq:1.8}))}\\
&=&\mbox{\boldmath $\gamma$}^\prime\xb\lb\Var(\by)(\xb\lb)^\prime\mbox{\boldmath $\gamma$}\\
&=&\sigma^2\mbox{\boldmath $\gamma$}^\prime(\xb\lb)(\xb\lb)^\prime\mbox{\boldmath $\gamma$}\qquad\mbox{(from 
(\ref{eq:1.5}))}\\
&=&\sigma^2\mbox{\boldmath $\gamma$}^\prime\{\pf+(\xb\lb-\pf)\}\{\pf+(\xb\lb-
\pf)\}^\prime\mbox{\boldmath $\gamma$}\\
&=&\sigma^2\mbox{\boldmath $\gamma$}^\prime\pf\mbox{\boldmath $\gamma$}+\sigma^2\mbox{\boldmath $\gamma$}^\prime(\xb\lb-\pf)(\xb\lb-
\pf)^\prime\mbox{\boldmath $\gamma$}\;,
\eeq
because $\xb\lb$ and $\pf$ are both matrices of projections on $\sr(\xb)$, 
and therefore \[(\xb\lb-\pf)\pf=\ze_{n,n}=\pf(\xb\lb-\pf)^\prime\;.\]
The variance of $\mbox{\boldmath $\eta$}^\prime \lb\by$ has been split into two components; the first 
is  \[\sigma^2\mbox{\boldmath $\gamma$}^\prime\pf\mbox{\boldmath $\gamma$}=\sigma^2\mbox{\boldmath $\eta$}^\prime\gb\mbox{\boldmath $\eta$}\;,\]
which is the variance of the least squares estimator. The second component is 
\[\sigma^2\mbox{\boldmath $\eta$}^\prime(\lb-\gb\xbp)(\lb-\gb\xbp)^\prime\mbox{\boldmath $\eta$}\;,\]
which by Theorem A.16 is non-negative and equal to zero when 
$\lb=\gb\xb^\prime$, i.e.\ $\mbox{\boldmath $\eta$}^\prime\lb\by$ is the least squares estimator. 

\bigskip
\noindent{\bf 1.6 Hypothesis testing}\\
The sums of squares in the analysis of variance (Table \ref{table:1.2}) take the general 
form $\by^\prime\pf^*\by$, where $\pf^*$ is the matrix of an orthogonal 
projection with $\rank(\pf^*)=q^*$, say. This section is concerned with some 
properties of the quadratic form $\by^\prime\pf^*\by$ which lead to a test of 
the general hypothesis 
\begin{equation}
\mbox{\boldmath $\pi$}=\ze_p\;.\hfill\
\label{eq:1.9}
\end {equation}

\noindent{\it Theorem 1.16.} If $\pf^*$ is a symmetric idempotent matrix of rank $q^*$ 
then
\begin{equation} 
\E[\by^\prime\pf^*\by]=\mbox{\boldmath $\pi$}^\prime\xbp\pf^*\xb\mbox{\boldmath $\pi$}+q^*\sigma^2\;.\hfill
\label{eq:1.10}
\end{equation}

\noindent{\it Proof.} \beq
&&\E[\by^\prime\pf^*\by]\\
&=&\E[\trace(\by^\prime\pf^*\by)]\\
&=&\E[\trace(\pf^*\by\by^\prime)]\qquad\mbox{(by Theorem A.10)}\\
&=&\trace(\pf^*\E[\by\by^\prime])\qquad\mbox{(by Theorem 1.9)}\\
&=&\trace\{\pf^*\E[(\xb\mbox{\boldmath $\pi$}+\mbox{\boldmath $\epsilon$})(\xb\mbox{\boldmath $\pi$}+\mbox{\boldmath $\epsilon$})^\prime]\}\\
&=&\trace\{\pf^*(\xb\mbox{\boldmath $\pi$}\mbox{\boldmath $\pi$}^\prime\xbp+\xb\mbox{\boldmath $\pi$}\E[\mbox{\boldmath $\epsilon$}^\prime]\\
&&\qquad\qquad\qquad+\E[\mbox{\boldmath $\epsilon$}]\mbox{\boldmath $\pi$}^\prime\xb^\prime+\E[\mbox{\boldmath $\epsilon$}\mbox{\boldmath $\epsilon$}^\prime])\}\\
&=&\trace(\pf^*\xb\mbox{\boldmath $\pi$}\mbox{\boldmath $\pi$}^\prime\xb^\prime+\sigma^2\pf^*)\;,
\eeq
since from (\ref{eq:1.6})
\[\E[\mbox{\boldmath $\epsilon$}]=\ze_n\qquad{\rm and}\qquad\E[\mbox{\boldmath $\epsilon$}\mbox{\boldmath $\epsilon$}^\prime]=\sigma^2\ib_n\;.\]
Now $\pf^*$ is an symmetric idempotent matrix therefore from Theorem A.37,
\[\trace(\pf^*)=q^*\;.\]
Hence 
\[\E[\by^\prime\pf^*\by]=\mbox{\boldmath $\pi$}^\prime\xb^\prime\pf^*\xb\mbox{\boldmath $\pi$}+q^*\sigma^2\;.\]
The expectation of the model sum of squares in Table \ref{table:1.2} is obtained by 
substituting $\pf^*=\pf$ and $q^*=q$ in (\ref{eq:1.10}), i.e.\ 
\begin {equation} 
\E[\by^\prime\pf^*\by]
=\mbox{\boldmath $\pi$}^\prime\xb^\prime\xb\mbox{\boldmath $\pi$}+q\sigma^2\;,\hfill\\
\label{eq:1.11}
\end {equation}
\beq
&&\qquad\qquad\qquad\mbox{since $\xbp\pf\xb=\xbp\xb$}\;.
\eeq
Similarly setting $\pf^*=(\ib_n-\pf)$ and $q^*=n-q$ in (\ref{eq:1.10}), the expectation 
of the residual sum of squares is
 \begin {equation} 
\E[\by^\prime(\ib_n-\pf)\by]
=(n-q)\sigma^2\;,\hfill\\
\label{eq:1.12}
\end {equation}
\beq
&&\qquad\qquad\qquad\mbox{since $(\ib_n-\pf)\xb=\ze_{n,p}$}\;.
\eeq
The matrix $(\ib_n-\pf)$ is called a sweep operator.

Thus the following result has been established:

\noindent{\it Theorem 1.17.} The residual mean square
\begin {equation}
s^2=\frac{1}{n-q} \by^\prime(\ib_n-\pf)\by\hfill\
\label{eq:1.13}
\end {equation}
 is an unbiased estimator of $\sigma^2$, i.e.
\[\E[s^2]=\sigma^2\;.\]
Under the hypothesis (\ref{eq:1.9}), the model mean square $\frac{1}{q}\by^\prime\pf\by$ 
is also an unbiased estimator of $\sigma^2$; consequently the ratio of the 
model and residual mean squares, known as the variance ratio, is parameter free 
and should be near one. In fact using the normality assumption (\ref{eq:1.7}) (which has 
not been needed so far) the theoretical distribution of the variance ratio can 
be derived, thereby providing a statistical test of the hypothesis (\ref{eq:1.9}). The 
remainder of this section will be concerned with this test.

The following results on distributions are stated but not proved. A detailed 
discussion is given by Searle (1971, Chapter~2.5).

\noindent{\it Theorem 1.18.} If $\by$ is distributed as $N(\xb\mbox{\boldmath $\pi$},\sigma^2\ib_n)$ 
then $\frac{1}{\sigma^2}\by^\prime\pf^*\by$ has a non-central $\chi^2$ 
distribution with $q^*$ degrees of freedom and non-centrality parameter 
$\frac{1}{2}\mbox{\boldmath $\pi$}^\prime\xbp\pf^*\xb\mbox{\boldmath $\pi$}$, if and only if $\pf^*$ is idempotent 
of rank $q^*$.

The following theorem, proved for the central case by Cochran (1934) and more 
generally by Madow (1940), is known as Cochran's Theorem. It establishes the 
independence of the quadratic forms $\by^\prime\pf^*\by$ in the analysis of 
variance table.
 
\noindent{\it Theorem 1.19.} If $\by$ is distributed as $N(\xb\mbox{\boldmath $\pi$}, \sigma^2\ib_n)$ 
and 
\[\by^\prime\by=\sum_{i=1}^l\by^\prime\pf_i^*\by\;,\]
where $\rank(\pf_i^*)=q_i^*$, \ ($i=1,\ldots,l$) then the quadratic forms 
$\frac{1}{\sigma^2}\by^\prime\pf_i^*\by$ have independent $\chi^2$ 
distributions with $q_i^*$ degrees of freedom if and only if \[\sum_{i=1}^l 
q_i^*=n\;.\]
Application of the preceding theorems to Table \ref{table:1.2}, together with (\ref{eq:1.11}) and 
(\ref{eq:1.12}) show that:

\noindent(i) $\frac{1}{\sigma^2}\by^\prime\pf\by$ has a non-central $\chi^2$ 
distribution with $q$ degrees of freedom and non-centrality parameter 
$\frac{1}{2}\mbox{\boldmath $\pi$}^\prime\xbp\xb\mbox{\boldmath $\pi$}$.

\noindent(ii) $\frac{1}{\sigma^2}\by^\prime(\ib_n-\pf)\by$ has a central 
$\chi^2$ distribution with $n-q$ degrees of freedom.

\noindent (iii) The model and residual sum of squares are independently 
distributed. The variance ratio 
\begin {equation}
\frac{\by^\prime\pf\by}{qs^2}\;,\hfill\
\label{eq:1.14}
\end {equation}
therefore has a non-central $F$ distribution with $q$ and $n-q$ degrees of 
freedom and non-centrality parameter $\frac{1}{2}\mbox{\boldmath $\pi$}^\prime\xbp\xb\mbox{\boldmath $\pi$}$.
Under the hypothesis (\ref{eq:1.9}), the statistic (\ref{eq:1.14}) has a central $F$ 
distribution, written as $F(q,n-q)$. The hypothesis can then be tested against 
the alternative that at least one of the elements of $\mbox{\boldmath $\pi$}$ is non-zero, using 
tables for the upper percentile points of $F(q,n-q)$. 

\bigskip
\noindent{\bf 1.7 Partitioning the model subspace}\\
The preceding theory has been general for any linear model of the form (\ref{eq:1.1}). 
In this section, attention will be directed specifically to the linear model 
for an experimental design with a number, say $d$, of factors named 
$F_1,\ldots,F_d$. The model (\ref{eq:1.1}) can then be partitioned in the following way:
\beq
\by
&=&\xb\mbox{\boldmath $\pi$}+\mbox{\boldmath $\epsilon$}\\
&=&\left[\begin{array}{cccc}
{\bf 1}_n&\xb_1&\ldots&\xb_d
\end{array}\right]
\left[\begin{array}{c}
\mu\\ \mbox{\boldmath $\pi$}_1\\ \vdots\\ \mbox{\boldmath $\pi$}_d
\end{array}\right]
+\mbox{\boldmath $\epsilon$}\\
\eeq
\begin {equation}
={\bf 1}_n\mu+\sum_{i=1}^d\xb_i\mbox{\boldmath $\pi$}_i+\mbox{\boldmath $\epsilon$}\;,\hfill
\label{eq:1.15}
\end {equation}
where:
\begin{enumerate}
\item[(i)] $\mu$ is a parameter common to each unit in the experiment and 
therefore corresponds to the population mean.
\item[(ii)] $\mbox{\boldmath $\pi$}_i$ is a vector of parameters for factor $F_i$; if $F_i$ has 
$p_i$ levels then $\mbox{\boldmath $\pi$}_i$ is of length $p_i$. 
\item[(iii)] $\xb_i$ is an $n\times p_i$ matrix of zeros and ones such that if 
the $h$th unit of the experiment receives the $j$th level of $F_i$, then the 
$(h,j)$th element of $\xb_i$ is one; otherwise the elements of $\xb_i$ are 
zero. The matrix $\xb_i$ is called the design matrix of factor $F_i$, \ 
($i=1,\ldots,d$).
\end{enumerate}

There are two major reasons why it is desirable to partition the model 
subspace, rather than work with $\mbox{\boldmath $\pi$}$ an the associated orthogonal projection 
matrix on $\sr(\xb)$. The first is that for the purpose of testing hypotheses 
about specific factors, it is necessary to partition the model sum of squares 
in the analysis of variance (Table \ref{table:1.2}) into components for each of the 
factors. The second reason is that often only the least squares estimators of a 
subset of $\mbox{\boldmath $\pi$}_1,\ldots,\mbox{\boldmath $\pi$}_d$ are needed. Therefore the remaining parameter  
vectors can be eliminated from the normal equations. 

These operations can effectively be handled by working with the orthogonal 
projection matrices 
\begin {equation}
\pf_i=\xb_i\gb_i\xbp_i\hfill\
\label{eq:1.16}
\end {equation}
on the subspaces $\sr(\xb_i)$, where $\gb_i$ is a generalized inverse of 
$\xbp_i\xb_i$, \ ($i=1,\ldots,d$) and the orthogonal projection matrix
\begin {equation}
\pf_G=\frac{1}{n}\jb_n\hfill\
\label{eq:1.17}
\end {equation}
on $\sr({\bf 1}_n)$ and $jb_n$ is an $n \times n$ matrix of ones.

Details for linear models with two or three factors are given in later 
sections. Firstly the concept of marginality is introduced; this allows $\mu$ 
to be eliminated from (\ref{eq:1.15}).

\bigskip
\noindent {\bf 1.8 Marginality}\\
In the linear model (\ref{eq:1.15}) it will be assumed that each of the $n$ units in the 
experimental design is assigned exactly one of the levels of each of the 
factors present in the model. In other words, in each row of the $\xb_i$, there 
is exactly one unit element and hence 
\begin {equation}
\xb_i{\bf 1}_{p_i}={\bf 1}_n\;,\hfill\
\label{eq:1.18}
\end {equation}
or alternatively
\begin {equation}
\sr({\bf 1}_n)
\subset\sr(\xb_i)\;,\hfill\\
\qquad\qquad(i=1,\ldots,d)\;.
\label{eq:1.19}
\end {equation}
The term for the population mean is said to be marginal to the factors $F_i$, \ 
($i=1,\ldots,d$). For the majority of common designs, the 
marginality condition (\ref{eq:1.19}) is automatically satisfied, however for 
more complex designs such as change-over designs it is necessary to introduce dummy or pseudo-factors 
in order to satisfy (\ref{eq:1.19}).

The following result can now be obtained directly from Theorem A.25.

\noindent{\it Theorem 1.20.} Two factors, with associated orthogonal projection matrices 
$\pf_1$ and $\pf_2$ are orthogonal if and only if 
\[\pf_1\pf_2=\pf_2\pf_1=\pf_G\;.\]
The concept of marginality can more generally be applied to factors. A detailed 
discussion is given by Nelder (1977). 

\noindent{\it Definition 1.6.} A factor $F_1$ is marginal to another factor 
$F_2$ if 
\[\sr(\xb_1)\subset\sr(\xb_2)\;,\]
where $\xb_i$ is the design matrix of $F_i$, \ ($i=1,2$). Let $\pf_1$ and 
$\pf_2$ be the orthogonal projection matrices of factors $F_1$ and $F_2$ 
respectively.
The following result is a direct consequence of Theorem A.24.

\noindent{\it Theorem 1.21.} If factor $F_1$ is marginal to factor $F_2$ then the orthogonal 
projection matrix on $\sr(\xb_2)\ominus\sr(\xb_1)$ is $\pf_2-\pf_1$. 

\noindent{\it Theorem 1.22.} If factor $F_1$ is marginal to factor $F_2$ then 
\[\pf_1\pf_2=\pf_2\pf_1=\pf_1\;.\]

\noindent{\it Proof.} From the previous theorem, the orthogonal subspaces 
$\sr(\xb_1)$ and $\sr(\xb_2)\ominus\sr(\xb_1)$ have orthogonal projection 
matrices $\pf_1$ and $\pf_2-\pf_1$ respectively. Then from Theorem A.23, 
\[\pf_1(\pf_2-\pf_1)=(\pf_2-\pf_1)\pf_1=\ze_{n,n}\;,\]
i.e.\
\[\pf_1\pf_2=\pf_2\pf_1=\pf_1\;.\]
\noindent{\it Theorem 1.23.} If factor $F_1$ is marginal to factor $F_2$ than a generalized inverse 
$\gb_2$ of $\xbp_2\xb_2$ is also a generalized inverse of $\xbp_2(\ib_n-
\pf_1)\xb_2$.

\noindent{\it Proof.} \beq
&&\xbp_2(\ib_n-\pf_1)\xb_2\gb_2\xbp_2(\ib_n-\pf_1)\xb_2\\
&=&(\xbp_2\xb_2)-\xbp_2\pf_1\xb_2)\gb_2(\xbp_2\xb_2-\xbp_2\pf_1\xb_2)\\
&=&\xbp_2\xb_2-2\xbp_2\pf_1\xb_2+\xbp_2\pf_1\pf_2\pf_1\xb_2\\
&&\qquad\mbox{(using Theorems 1.3 and 1.4)}\;,\\
&=&\xbp_2\xb_2-\xbp_2\pf_1\xb_2\qquad\mbox{(from Theorem 1.22)}\;,\\
&=&\xbp_2(\ib_n-\pf_1)\xb_2\;.
\eeq
Hence $\gb_2$ is a generalized inverse of $\xbp_2(\ib_n-\pf_1)\xb_2$.

In addition to the above theorems for two factors, analogous results for the 
relationship between the population mean and the factors $F_i$ are summarized 
in the following theorem:

\noindent{\it Theorem 1.24.} For a factor $F_i$ with design matrix $\xb_i$ and orthogonal projection 
matrix $\pf_i$, 
\begin{enumerate}
\item[(i)] the orthogonal projection matrix on 
\[\sr(\xb_i)\ominus\sr({\bf 1}_n)\quad{\rm is}\quad\pf_i-\pf_G\;,\]
\item[(ii)] $\qquad\pf_i\pf_G=\pf_G\pf_i=\pf_G\;,$
\item[(iii)] a generalized inverse of $\xbp_i\xb_i$ is also a generalized 
inverse of $\xbp_i(\ib_n-\pf_G)\xb_i$, \ ($i=1,\ldots,d$).
\end{enumerate}

\bigskip
\noindent{\bf 1.9 Eliminating the mean}\\
This section is concerned with simplifying the linear model (\ref{eq:1.15}) by removing 
the term for the population mean. Assuming the marginality constraint (\ref{eq:1.19}), 
it is sufficient for the purposes of this section to consider a one factor 
model, i.e.\ $d=1$ in (\ref{eq:1.15}).
Then 
\begin {equation}
\xb=\left[\begin{array}{cc}
{\bf 1}_n&\xb_1\end{array}\right]\;,\hfill\
\label{eq:1.20}
\end {equation}
and from (\ref{eq:1.3}), the normal equations become 
\[\left[\begin{array}{cc}
n&{\bf 1}_n^\prime\xb_1\\
\xbp_1{\bf 1}_n&\xbp_1\xb_1
\end{array}\right]
\left[\begin{array}{c}
\hat\mu\\ \hat{\mbox{\boldmath $\pi$}}_1
\end{array} \right]
=\left[\begin{array}{c}
{\bf 1}^\prime_n\\\xbp_1
\end{array}\right]
\by\;.\]
The only estimable functions involve comparisons between the elements of 
$\mbox{\boldmath $\pi$}_1$, so $\hat\mu$ can be eliminated from the above equations. For example
\[\hat\mu=\frac{1}{n}({\bf 1}^\prime_n\by-{\bf 1}_n^\prime\xb_1\hat{\mbox{\boldmath $\pi$}}_1)\;,\]
and 
\[{\bf 1}_n\xb_1\hat\mu+\xbp_1\xb_1\hat{\mbox{\boldmath $\pi$}}_1=\xbp_1\by\;.\]
Substitution then gives
\begin {equation}
\xbp_1(\ib_n-\pf_G)\xb_1\hat{\mbox{\boldmath $\pi$}}_1=\xbp_1(\ib_n-\pf_G)\by\;.\hfill\
\label{eq:1.21}
\end {equation}
Therefore using Theorem 1.24, the least squares estimator of an estimable 
linear function $\mbox{\boldmath $\eta$}^\prime\hat{\mbox{\boldmath $\pi$}}_1$ is given by 
\begin {equation}
\mbox{\boldmath $\eta$}^\prime\hat{\mbox{\boldmath $\pi$}}_1=\mbox{\boldmath $\eta$}^\prime\gb_1\xbp_1\by^*\;,\hfill\
\label{eq:1.22}
\end {equation}
where $\by^*=(\ib_n-\pf_G)\by$ is a vector of mean corrected observations obtained by applying the sweep operator for the mean,
namely $(\ib_n-\pf_G)$. 

Condition (\ref{eq:1.19}) means that the model subspace can be partitioned as the direct 
sum 
\[\sr(\xb)=\sr({\bf 1}_n)\oplus\{\sr(\xb_1)\ominus\sr({\bf 1}_n)\}\;.\]
From Theorem 1.24 the orthogonal projection matrix on 
$\sr(\xb_1)\ominus\sr({\bf1}_n)$ is
\[\pf_1-\pf_G=\pf_1(\ib_n-\pf_G)\;.\]
The model sum of squares in Table \ref{table:1.2} can therefore be partitioned into a 
component $\by^\prime\pf_G\by$ with 1 degree of freedom for the mean of the 
observations and a component $\by^\prime(\pf_1-
\pf_G)\by=\by^{*^\prime}\pf_1\by^*$ with $q_1=q-1$ degrees of freedom for 
testing the hypothesis 
\[\mbox{\boldmath $\pi$}_1=\ze_{p_1}\;.\]
It is usual to subtract the correction factor $\by^\prime\pf_G\by$ from the 
total sum of squares $\by^\prime\by$ in the analysis of variance in Table \ref{table:1.2}, 
to give the mean corrected total sum of squares $\by^\prime(\ib_n-
\pf_G)\by=\by^{*^\prime}\by^*$ on $n-1$ degrees of freedom. A more common form 
for the analysis of variance is given in Table \ref{table:1.3}.

\begin {table}
\centering\small
\caption{{\it Analysis of Variance}}
\vspace{0.2cm}
\begin{tabular}{lcc}
\hline
&{\bf df}&{\bf ss}\\
\multicolumn{1}{c}{\it Units stratum}&&\\
$F_1$&$q_1$&$\by^{*^\prime}\pf_1\by^*$\\
Residual&$n-q_1-1$&$\by^{*^\prime}(\ib_n-\pf_1)\by^*$\\
\hline
Grand Total&$n-1$&$\by^{*^\prime}\by^*$\\
\hline
\end{tabular}
\label{table:1.3}
\end{table}

The least squares estimator (\ref{eq:1.22}) and the analysis of variance (Table \ref{table:1.3}) 
show that one can write the linear model for one factor in the equivalent form 
\begin {equation}
\by^*=\xb_1\mbox{\boldmath $\pi$}_1+\mbox{\boldmath $\epsilon$}\;.\hfill\
\label{eq:1.23}
\end {equation}
Similarly with condition (\ref{eq:1.19}), the linear model (\ref{eq:1.15}) for $d$ factors can be 
written as 
\beq
\by^*
&=&\xb\mbox{\boldmath $\pi$}+\mbox{\boldmath $\epsilon$}\\
\eeq
\begin {equation}
=\sum_{i=1}^d\xb_i\mbox{\boldmath $\pi$}_i+\mbox{\boldmath $\epsilon$}\;.\hfill
\label{eq:1.24}
\end {equation}
\bigskip
\noindent{\bf 1.10 Two-factor model}\\
In this section the linear model for two non-orthogonal factors $F_1$ and $F_2$ 
is considered in detail. From (\ref{eq:1.24}) the model for $d=2$ can be written in the 
form 
\begin {equation}
\by^*=\xb_1\mbox{\boldmath $\pi$}_1+\xb_2\mbox{\boldmath $\pi$}_2+\mbox{\boldmath $\epsilon$}\;.\hfill\
\label{eq:1.25}
\end {equation}
The normal equations for the least squares estimation of $\mbox{\boldmath $\pi$}_1$ and $\mbox{\boldmath $\pi$}_2$ 
are 
\begin {equation}
\left[\begin{array}{cc}
\xbp_1\xb_1&\xb_1^\prime\xb_2\\
\xbp_2\xb_1&\xbp_2\xb_2
\end{array}\right]
\left[\begin{array}{c}
\hat{\mbox{\boldmath $\pi$}}_1\\\hat{\mbox{\boldmath $\pi$}}_2
\end{array}\right]
=\left[\begin{array}{c}
\xbp_1\\\xbp_2
\end{array}\right]\by^*\;.\hfill\
\label{eq:1.26}
\end {equation}
If one is mainly interested in say the estimation of the parameter vector 
$\mbox{\boldmath $\pi$}_2$ for factor $F_2$, then $\hat{\mbox{\boldmath $\pi$}}_1$ can be eliminated from the normal 
equations to give what are known as reduced normal equations.

\noindent{\it Theorem 1.25.} The reduced normal equations for the estimation of $\mbox{\boldmath $\pi$}_2$ are given by
 \begin {equation} 
\tilde\xbp_2\tilde\xbp_2\hat{\mbox{\boldmath $\pi$}}_2
=\tilde\xbp_2\by^*\;,\\
\label{eq:1.27}
\end {equation}
\beq
{\rm where} &&\tilde\xb_2=(\ib_n-\pf_1)\xb_2\;.
\eeq
\noindent{\it Proof.}
Let $\gb_1$ be a generalized inverse of $\xb_1^\prime\xb_1$. From (\ref{eq:1.26}) the 
normal equations are the two sets
\begin {equation}
\begin{array}{l}
\xbp_1\xb_1\hat{\mbox{\boldmath $\pi$}}_1+\xbp_1\xb_2\hat{\mbox{\boldmath $\pi$}}_2=\xbp_1\by^*\\
\xbp_2\xb_1\hat{\mbox{\boldmath $\pi$}}_1+\xbp_2\xb_2\hat{\mbox{\boldmath $\pi$}}_2=\xbp_2\by^*\;,
\end{array}\hfill\
\label{eq:1.28}
\end {equation}
whence $\hat{\mbox{\boldmath $\pi$}}_1$ can be eliminated by premultiplying the first set by 
$\xbp_2\xb_1\gb_1$ and subtracting from the second set. This operation gives
\[(\xbp_2\xb_2-\xbp_2\xb_1\gb_1\xbp_1\xb_2)\hat{\mbox{\boldmath $\pi$}}_2=\xbp_2\by^*-
\xbp_2\xb_1\gb_1\xbp_1\by^*\;,\]
i.e.\ $\qquad\xbp_2(\ib_n-\pf_1)\xb_2\hat{\mbox{\boldmath $\pi$}}_2=\xbp_2(\ib_n-\pf_1)\by^*\;,$
where $\pf_1=\xb_1\gb_1\xbp_1$, and therefore
\[\tilde\xbp_2\tilde\xb_2\hat{\mbox{\boldmath $\pi$}}_2=\tilde\xb_2^\prime\by^*\;.\]
The solution of (\ref{eq:1.27}) gives the least squares estimator of the parameter 
vector for factor $F_2$, adjusted for the effect of $F_1$. If the model had 
been 
\[
\by^*=\xb_2\mbox{\boldmath $\pi$}_2+\mbox{\boldmath $\epsilon$}\;,\]
then the solution of the normal equations
\begin {equation}
\xbp_2\xb_2\hat{\mbox{\boldmath $\pi$}}_2=\xbp_2\by^*\;,\hfill\
\label{eq:1.29}
\end {equation}
would have given the least squares estimator, ignoring the effect of $F_1$. 

It is interesting to note that the form of the equations (\ref{eq:1.27}) and (\ref{eq:1.29}) is 
the same; the difference being that when it is required to adjust for $F_1$, \ 
$\xb_2$ is replaced by $(\ib_n-\pf_1)\xb_2$. When factors $F_1$ and $F_2$ are 
orthogonal then the solutions of the two sets of equations are the same. This 
is because from Theorem 1.20
\begin {equation}
\xbp_2\pf_1=\xb_2^\prime\pf_2\pf_1=\xbp_2\pf_G\;.\hfill\
\label{eq:1.30}
\end {equation}
The sum of squares for $F_2$ adjusted for the effect of $F_1$ is
\beq \hat{\mbox{\boldmath $\pi$}}^\prime_2\tilde\xbp_2\by^*
&=&\by^{*^\prime}\tilde\pf_2\by^*\;,\\
&{\rm where}&\tilde\pf_2=\tilde\xb_2\tilde\gb_2\tilde\xbp_2\;,
\eeq
and $\tilde\gb_2$ is a generalized inverse of $\tilde\xbp_2\tilde\xb_2$. 
The following theorem shows how the model sum of squares 
$\hat{\mbox{\boldmath $\pi$}}^\prime\xbp\by^*$ can be partitioned into the sum of squares 
$\by^{*^\prime}\pf_1\by^*$ for $F_1$ ignoring $F_2$ and the sum of squares for 
$F_2$ adjusted for $F_1$. 

\noindent{\it Theorem 1.26.}
For the model (\ref{eq:1.25}), 
\[\hat{\mbox{\boldmath $\pi$}}^\prime\xbp\by^*=\by^{*^\prime}\pf_1\by^*+\by^{*^\prime}\tilde\pf_2\b
y^*\;.\]

\noindent{\it Proof.} Premultiplying the first set of (\ref{eq:1.28}) by $\xb_1\gb_1$ gives 
\[\xb_1\hat{\mbox{\boldmath $\pi$}}_1+\pf_1\xb_2\hat{\mbox{\boldmath $\pi$}}_2=\pf_1\by^*\;.\]
Therefore
\[\hat{\mbox{\boldmath $\pi$}}_1^\prime\xbp_1\by^*=\byp^*\pf_1\by^*-\hat{\mbox{\boldmath $\pi$}}_2^\prime\xbp_2\pf_1\by^*\;.\]
Now
\beq \hat{\mbox{\boldmath $\pi$}}^\prime\xb^\prime\by^*
&=&[\hat{\mbox{\boldmath $\pi$}}_1^\prime\hat{\mbox{\boldmath $\pi$}}_2^\prime]
\left[\begin{array}{c}
\xbp_1\\\xbp_2 \end{array}\right]\by^*\;,\\
&=&\hat{\mbox{\boldmath $\pi$}}_1^\prime\xb_1^\prime\by^*+\hat{\mbox{\boldmath $\pi$}}_2^\prime\xb^\prime_2\by^*\;,\\
&=&(\by^{*^\prime}\pf_1\by^*-
\hat{\mbox{\boldmath $\pi$}}_2^\prime\xb_2^\prime\pf_1\by^*)+\hat{\mbox{\boldmath $\pi$}}_2^\prime\xbp_2\by^*\;,\\
&=&\by^{*^\prime}\pf_1\by^*+\hat{\mbox{\boldmath $\pi$}}_2^\prime\xbp_2(\ib_n-\pf_1)\by^*\;,\\
&=&\by^{*^\prime}\pf_1\by^*+\by^{*^\prime}\tilde\pf_2\by^*\;.
\eeq
If rank $(\tilde\xb_2)=\tilde q_2$, then the analysis of variance table takes 
the form given in Table \ref{table:1.4}. The expression for the residual sum of squares is 
obtained by difference but it can be written in the alternate form 
\begin {equation}
\by^{*^\prime}(\ib_n-\pf_1-\tilde\pf_2)\by^*=\by^{\spr}(\ib_n-
\tilde\pf_2)(\ib_n-\pf_1)\by^*\;,\
\label{eq:1.31}
\end {equation}
since 
\[
\pf_2\pf_1=(\ib_n-\pf_1)\xb_2\tilde\gb_2\xbp_2(\ib_n-
\pf_1)\pf_1=\ze_{n,n}\;.\hfill\
\]

The expectation of the sum of squares for $F_1$ ignoring $F_2$ involves both 
$\mbox{\boldmath $\pi$}_1$ and $\mbox{\boldmath $\pi$}_2$; however the following theorem shows that the expectation 
of the sum of squares for $F_2$ adjusted for $F_1$ is independent of $\mbox{\boldmath $\pi$}_1$. 

\noindent{\it Theorem 1.27.} For the model (\ref{eq:1.25}), 
\begin {equation}
\E[\by^{\spr}\pf_2\by^*]=\mbox{\boldmath $\pi$}_2^\prime\tilde\xbp_2\tilde\xb_2\mbox{\boldmath $\pi$}_2+\tilde 
q_2\sigma^2\;.\hfill\
\label{eq:1.32}
\end {equation}

\noindent{\it Proof.} By substituting $\pf^*=\tilde\pf_2$ and $q^*=\tilde q_2$ 
in Theorem 1.16, 
\[\E[\by^{\spr}\tilde\pf_2\by^*]=(\xb_1\mbox{\boldmath $\pi$}_1+\xb_2\mbox{\boldmath $\pi$}_1)^\prime\tilde\pf_2(\xb_
1\mbox{\boldmath $\pi$}_1+\xb_2\mbox{\boldmath $\pi$}_2)+\tilde q_2\sigma^2\;.\]
Now as in (\ref{eq:1.31})
\[\tilde \pf_2\xb_1=\ze_{n,p_1}\;,\] and so
\beq \E[\by^{\spr}\tilde\pf_2\by^*]
&=&\mbox{\boldmath $\pi$}_2^\prime\tilde\xbp_2\tilde\xb_2\tilde\gb_2\tilde\xbp_2\tilde\xb_2\mbox{\boldmath $\pi$}_2
+\tilde q_2\sigma^2\\
&=&\mbox{\boldmath $\pi$}_2^\prime\tilde\xbp_2\tilde\xb_2\mbox{\boldmath $\pi$}_2+\tilde q_2\sigma^2\;.
\eeq

From Theorems 1.18 and 1.19, the quadratic forms in Table \ref{table:1.4} have independent 
$\chi^2$ distributions. Therefore under the hypothesis \[\mbox{\boldmath $\pi$}_2=\ze_{p_2}\;,\]
the variance ratio \[\frac{\by^{\spr}\tilde\pf_2\by^*}{\tilde q_2s^2}\]
has an $F(\tilde q_2,n-q_1-\tilde q_2-1)$ distribution.

\begin{table}
\centering\small
\caption{{\it Analysis of Variance}}
\vspace{0.2cm}
\begin{tabular}{lcc}
\hline
&{\bf df}&{\bf ss}\\
\multicolumn{1}{c}{\sl Units stratum}&&\\
$F_1$ (ig.\ $F_2$)&$q_1$&$\by^{*^\prime}\pf_1\by^*$\\
$F_2$ (adj.\ $F_1$)&$\tilde q_2$&$\by^{*^\prime}\tilde\pf_2\by^*$\\
Residual&$n-q_1-\tilde q_2-1$&$\by^{*^\prime}(\ib_n-\pf_1-\tilde\pf_2)\by^*$\\
\hline
Grand Total&$n-1$&$\by^{*^\prime}\by^*$\\
\hline
\end{tabular}
\label{table:1.4}
\end{table}
Clearly $\hat{\mbox{\boldmath $\pi$}}^\prime\xbp\by$ can also be partitioned as the sum of squares 
for $F_1$ adjusted for $F_2$ and the sum of squares for $F_2$ ignoring $F_1$. 
When $F_1$ is marginal 
to $F_2$, it follows directly from Theorem 1.21, or with calculation using 
Theorems 1.22 and 1.23 that $\tilde\pf_2=\pf_2-\pf_1$; also from Theorem A.20, 
$\tilde q_2=q_2-q_1$ where $q_2=\rank(\xb_2)$. The analysis of variance then 
takes the simpler form given in Table \ref{table:1.5}.
\begin{table}
\centering\small
\caption{{\it Analysis of Variance}}
\vspace{0.2cm}
\begin{tabular}{lcc}
\hline
&{\bf df}&{\bf ss}\\
\multicolumn{1}{c}{\sl Units stratum}&&\\
$F_1$ (ig.\ $F_2$)&$q_1$&$\by^{*^\prime}\pf_1\by^*$\\
$F_2$ (adj.\ $F_1$)&$q_2-q_1$&$\by^{*^\prime}(\pf_2-\pf_1)\by^*$\\
Residual&$n-q_2-1$&$\by^{*^\prime}(\ib_n-\pf_2)\by^*$\\
\hline
Grand Total&$n-1$&$\by^{*^\prime}\by^*$\\
\hline
\end{tabular}
\label{table:1.5}
\end{table}

\bigskip
\noindent {\bf 1.11 Three-factor model}\\
For the linear model
\[\by^*=\xb_1\mbox{\boldmath $\pi$}_1+\xb_2\mbox{\boldmath $\pi$}_2+\xb_3\mbox{\boldmath $\pi$}_3+\mbox{\boldmath $\epsilon$}\;,\]
the normal equations are 
\[\left[\begin{array}{lll}
\xbp_1\xb_1&\xbp_1\xb_2&\xbp_1\xb_3\\
\xbp_2\xb_1&\xbp_2\xb_2&\xbp_2\xb_3\\
\xbp_3\xb_1&\xbp_3\xb_2&\xbp_3\xb_3
\end{array}\right]
\left[\begin{array}{c}
\hat{\mbox{\boldmath $\pi$}}_1\\
\hat{\mbox{\boldmath $\pi$}}_2\\
\hat{\mbox{\boldmath $\pi$}}_3
\end{array}\right]
=\left[\begin{array}{c}
\xbp_1\\
\xbp_2\\
\xbp_3
\end{array}\right]\by^*\;.\]
If the estimation of $\mbox{\boldmath $\pi$}_3$ is of primary interest then $\mbox{\boldmath $\pi$}_1$ and $\mbox{\boldmath $\pi$}_2$ 
can be eliminated from the above. The following result is proved along the 
lines used for Theorem 1.25.

\noindent{\it Theorem 1.28.} The reduced normal equations for the estimation of $\mbox{\boldmath $\pi$}_3$ are given by 
\[\tilde{\tilde\xbp}_3
\tilde{\tilde\xb}_3\hat{\mbox{\boldmath $\pi$}}_3=\tilde
{\tilde\xbp}_3\by^*\;,\]
where $\qquad \tilde{\tilde\xb}_3=(\ib_n-
\tilde\pf_2)\tilde\xb_3$,
\[\tilde\pf_2=\tilde\xb_2\tilde\gb_2\tilde\xbp_2\quad{\rm 
,}\quad\tilde\xb_i=(\ib_n-\pf_1)\xb_i\;,\quad i=2,3\]
and 
${\tilde\gb}_2$ is a generalized inverse of 
${\tilde\xbp}_2{\tilde\xb}_2$.

Define $\tilde{\tilde\pf}_3=
\tilde{\tilde\xb}_3
\tilde{\tilde\gb}_3
\tilde{\tilde\xbp}_3$ where 
$\tilde{\tilde\gb}_3$ is a generalized inverse of 
$\tilde{\tilde\xbp}_3
\tilde{\tilde\xb}_3$ and let $\rank(
\tilde{\tilde\xb}_3)=\tilde{\tilde 
q}_3$. Then the analysis of variance (Table \ref{table:1.6}) for three non-orthogonal 
factors can be written as a straightforward extension of Table \ref{table:1.4}.

\begin{table}
\centering\small
\caption{{\it Analysis of Variance}}
\vspace{0.2cm}
\begin{tabular}{lcc}
\hline
&{\bf df}&{\bf ss}\\
\multicolumn{1}{c}{\sl Units stratum}&&\\
$F_1$ (ig.\ $F_2$ and $F_3$)&$q_1$&$\by^{*^\prime}\pf_1\by^*$\\
$F_2$ (adj.\ $F_1$, ig.\ $F_3$)&$\tilde q_2$&$\by^{*^\prime}\tilde\pf_2\by^*$\\
$F_3$ (adj.\ $F_1$, and $F_2$)&$\tilde{\tilde q}_3$
&$\by^{*^\prime}\tilde{\tilde\pf}_3\by^*$\\
Residual&$n-q_1-\tilde q_2-\tilde{\tilde q}_3-1$
&$\by^{*^\prime}(\ib_n-\pf_1-\tilde\pf_2-
\tilde{\tilde\pf}_3)\by^*$\\
\hline
Grand Total&$n-1$&$\by^{*^\prime}\by^*$\\
\hline
\end{tabular}
\label{table:1.6}
\end{table}
The residual sum of squares in Table \ref{table:1.6} can be written in the alternate form
\begin {equation}
\by^{\spr}(\ib_n-\pf_1-\tilde\pf_2-
\tilde{\tilde\pf}_3)\by^*=\by^{\spr}(\ib_n-
\tilde{\tilde\pf}_3)(\ib_n-\tilde\pf_2)(\ib_n-
\pf_1)\by^*\;.\hfill\
\label{eq:1.33}
\end {equation}
If $F_1$ and $F_2$ are marginal to $F_3$ then $\tilde{\tilde\pf}_3
=\pf_3-\tilde\pf_2-\pf_1$ and $\tilde{\tilde q}_3=q_3-
\tilde q_2-q_1$, where $q_3=\rank (\xb_3)$. The analysis of variance then takes 
the form given in Table \ref{table:1.7}.

\begin{table}
\centering\small
\caption{{\it Analysis of Variance}}
\vspace{0.2cm}
\begin{tabular}{lcc}
\hline
&{\bf df}&{\bf ss}\\
\multicolumn{1}{c}{\sl Units stratum}&&\\
$F_1$ (ig.\ $F_2$ and $F_3$)&$q_1$&$\by^{*^\prime}\pf_1\by^*$\\
$F_2$ (adj.\ $F_1$, ig.\ $F_3$)&$\tilde q_2$&$\by^{*^\prime}\tilde\pf_2\by^*$\\
$F_3$ (adj.\ $F_1$, and $F_2$)&$q_3-\tilde q_2-q_1$&$\by^{*^\prime}(\pf_3-\tilde\pf_2
-\pf_1)\by^*$\\
Residual&$n-q_3-1$
&$\by^{*^\prime}(\ib_n-\pf_3)\by^*$\\
\hline
Grand Total&$n-1$&$\by^{*^\prime}\by^*$\\
\hline
\end{tabular}
\label{table:1.7}
\end{table}
If in addition $F_1$ and $F_2$ are orthogonal factors then $\tilde \pf_2=\pf_2-
\pf_G$ and $\tilde q_2=q_2$. Therefore $\tilde \pf_2$ and $\tilde q_2$ in Table \ref{table:1.7} can be replaced by $\pf_2$ and $q_2$ respectively. 

\newpage

\begin{center}
{\bf 2. Block Designs}
\end{center}
\bigskip
\noindent {\bf 2.1 The model}\\
Section 1.10 gave a general presentation of the two-factor model.  A particular case is where the first factor is included to account for extraneous effects such as blocks in a field trial; the levels of the second factor are the items of interest.  It is usual to refer to these factors as 'blocks' and 'treatments' respectively.  In this section we consider a block design with $b$ blocks and $v$ treatments.  The general terminology of Section 1.10 is made more specific to the block design model and (\ref{eq:1.25}) is rewritten as
\begin {equation}
\bys={\bf Z}\mbox{\boldmath $\beta$}+ {\bf X}\mbox{\boldmath $\tau$}+\mbox{\boldmath $\epsilon$},
\label{eq:2.34}
\end {equation}
where $\bys$ is an $n \times 1$ vector of mean corrected observations; in other words the grand mean has been swept from the data (Section 1.9). The $n \times b$ matrix ${\bf Z}$ and $n \times v$ matrix ${\bf X}$ are the design matrices for blocks and treatments respectively (Section 1.7) and $\bfb$  and $\bft$ are $b \times 1$ and $v \times 1$ vectors of parameters for blocks and treatments respectively. 

The projection matrices (\ref{eq:1.16}) for blocks and treatments are written as ${\bf P}_B$ and ${\bf P}_T$, respectively.  For a randomized complete block design, blocks and treatments are orthogonal, i.e. ${\bf P}_B {\bf P}_T={\bf P}_G$, (Theorem 1.20) but for an incomplete block design, blocks need to be eliminated from the least squares equations to give the reduced normal equations  (Theorem 1.25) for the estimation of treatment effects; these are written as
\begin {equation} 
\tilde\xbp\tilde\xb\hat{\mbox{\boldmath $\tau$}}
=\tilde\xbp\bys\;,\\
\label{eq:2.35}
\end {equation}
\beq
{\rm where} &&\tilde\xb=(\ib_n-{\bf P}_B)\xb\;.
\eeq

In this section, attention will be restricted to incomplete block designs where all the treatments are replicated the same number of times ($r$) and all the blocks have the same number of units ($k$).  For such designs $\xbp\xb=rI_v$ and  $\zbp\zb=kI_b$.  Then

\begin {equation} 
\tilde\xbp\tilde\xb
=r\ib_v-\frac{1}{k}\nnp.\\
\label{eq:2.36}
\end {equation}
where ${\bf N} =\xbp\zb$ is the $v \times b$ incidence matrix of the block design. 
The $v \times v$ matrix $\nnp$ is called the called the concurrence matrix of the design; the $(i,i^\prime)$th element of $\nnp$ is the number of times treatments $i$ and $i^\prime$ appear together in the same block.  The sum of the elements in any row or column of $\nnp$ is equal to $rk$, i.e. $\nnp {\bf 1}_v=rk {\bf 1}_v$.  Hence $\tilde\xbp\tilde\xb\hat {\bf 1}_v= {\bf 0}_v$ and in the terminology of Appendix A, ${\bf 1}_v$ is in the kernel of $\tilde\xbp\tilde\xb$, i.e. $\ck(\tilde\xbp\tilde\xb)$.

Only connected designs are considered here; for such designs every pair of treatments can be linked through block either directly (i.e. in the same block) or indirectly (via other treatments).  For connected designs, ${\rm rank}(\tilde\xbp\tilde\xb)=v-1$ and $\ck(\tilde\xbp\tilde\xb)$ is in the subspace consisting of scalar multiples of ${\bf 1}_v$. Also as a consequence of the general theory in Section 1.2, it is possible to work ${\rm modulo}\  {\bf J}_v$ when constructing a generalized inverse ${\tilde\xbp\tilde\xb}^-$ of $\tilde\xbp\tilde\xb$; i.e. scalar multiples of ${\bf J}_v$ can be added to or subtracted from $\tilde\xbp\tilde\xb$ or ${\tilde\xbp\tilde\xb}^-$ to simplify calculations. In later sections some operations will be carried out ${\rm modulo}\  {\bf J}_v$.

The least squares solution $\hat{\mbox{\boldmath $\tau$}}$ is non-unique although any two solutions differ only by a scalar multiple of ${\bf 1}_v$; estimable functions of $\mbox{\boldmath $\tau$}$ are unique.  Often the solution $\hat{\mbox{\boldmath $\tau$}}$ where ${\bf 1}_v^\prime \hat{\mbox{\boldmath $\tau$}}=0$ is chosen; the elements of $\hat{\mbox{\boldmath $\tau$}}$ are then called the treatment effects.  Scalar multiples of ${\bf J}_v$ can be added on to the variance matrix of $\hat{\mbox{\boldmath $\tau$}}$, namely $\sigma^2 {\tilde\xbp\tilde\xb}^-$.  However the variance $\sigma^2 \bfe^\prime{\tilde\xbp\tilde\xb}^- \bfe$ of an estimable function of $\mbox{\boldmath $\tau$}$ is unique.  Often ${\tilde\xbp\tilde\xb}^-$ is chosen so that ${\tilde\xbp\tilde\xb}^- {\bf 1}_v={\bf 0}_v$.

\bigskip
\noindent {\bf 2.2  Analysis of variance}\\
The analysis of variance (Table \ref{table:2.8}) for the linear model (\ref{eq:2.34}) follows directly from Table \ref{table:1.4}.  The total sum of squares is partioned into three components.  The first is a term $\bysp {\bf P}_B \bys$ for the blocks sum of squares ignoring treatments, i.e. no adjustments are made to the block totals to take account of the fact that treatments are not orthogonal to blocks.  Hence the blocks sum of squares contains some information on treatments.  The second component is a term $\bysp \tilde{\bf P}_T \bys$ for the treatments sum of squares having adjusted for the fact that not all treatments appear in each block.  For a two-factor model such as  (\ref{eq:2.34}), Theorem 1.26 demonstrates how the model sum of squares can be partitioned.  The final component is the residual sum of squares, $\bysp (\ib_n-{\bf P}_B-\tilde{\bf P}_T) \bys$.  This quantity can be thought of as the sum of squares of the elements of the vector of residuals
\begin {equation}
(\ib_n-{\bf P}_B-\tilde{\bf P}_T) \bys.
\label{eq:2.37}
\end {equation}
The matrix $(\ib_n-{\bf P}_B-\tilde{\bf P}_T)$ which operates on $\bys$ to produce the vector of deviations from a least squares fit of model (\ref{eq:2.34}) is called the residual operator.  Using (\ref{eq:1.31}) the residual operator can be factorized into the product of two sweep operators, i.e.
\begin {equation}
(\ib_n-{\bf P}_B-\tilde{\bf P}_T) = (\ib_n-\tilde{\bf P}_T)(\ib_n-{\bf P}_B).
\label{eq:2.38}
\end {equation}

\begin{table}
\centering\small
\caption{{\it Analysis of Variance}}
\vspace{0.2cm}
\begin{tabular}{llcc}
\hline
{\bf Source} & &{\bf df}&{\bf ss}\\
{\it Blocks stratum}&&&\\
&Total&$b-1$&$\by^{*^\prime}\pf_B\by^*$\\
{\it Blocks.plots stratum} &&&\\ 
& Treatments (adj.)&$v-1$&$\by^{*^\prime}\tilde\pf_T\by^*$\\
& Residual&$n-v-b-1$&$\by^{*^\prime}(\ib_n-\pf_B-\tilde\pf_T)\by^*$\\
\hline
Grand Total&&$n-1$&$\by^{*^\prime}\by^*$\\
\hline
\end{tabular}
\label{table:2.8}
\end{table}

The complicated nature of the projection matrix $\tilde\pf_T$ reflects the allowance for non-orthogonality between blocks and treatments.  Whereas $\pf_T \bys$ (and $\pf_B \bys$) involve simple averaging operations on $\bys$, $\tilde\pf_T\bys$ in general requires the calculation of a generalized inverse of the matrix of the reduced normal equations, $\tilde\xbp\tilde\xb$.  However for some incomplete block designs the complicated sweep operator $(\ib_n-\tilde{\bf P}_T)$ can be factorized into a sequence of simpler operations.  A description of this process is left until Section 3.3.

\bigskip
\noindent {\bf 2.3  Efficiency factors}\\
In a randomized complete block design all the information available on treatment differences can be obtained from comparisons within blocks, i.e. there is no information on treatments in the {\it Blocks stratum}.  This is because all treatments appear in each block so any contrast $\bfe^\prime \hat\bft$ can be computed wholly within blocks. However in an incomplete block design, the information on treatment comparisons is partitioned into that contained in within-block comparisons and information which is incorporated into block totals.
For example in the incomplete block design in Table \ref{table:2.9}, consider the contrast $\bfe^\prime \hat\bft$ where  $\bfe^\prime =(1,-1,-1,1)$; this compares treatments 1 and 4 with treatments 2 and 3.  In the first two blocks the contrast is orthogonal to blocks because it involves the difference between treatments 1 and 3 in block 1 and treatments 2 and 4 in block 2.  Thus any block effects cancel out.  But in the other two blocks of the design, information on the contrast is totally bound up or confounded with block totals.  Therefore for the contrast of interest half of the information would be contained in the {\it Blocks.plots stratum} and the other half in the {\it Blocks stratum}.  In general, treatment information in an incomplete block design is spread over two or more strata.

\begin{table}
\centering\small
\caption{{\it An incomplete block design for 4 treatments in blocks of size 2}}
\vspace{0.2cm}
\begin{tabular}{lcccc}
{\bf Block} & 1 & 2 & 3 & 4\\
\cline{2-5}
& 1 & 2 & 2 & 1\\
& 3 & 4 & 3 & 4 \\
\end{tabular}
\label{table:2.9}
\end{table}

The least squares estimate $\hat\bft$ obtained from (\ref{eq:2.35}) only uses within-block information and is known as the intra-block treatment estimate. The concept of an efficiency factor is a measure of the proportion of information available on treatments in the intra-block analysis.  For the above example, the contrast  $\bfe^\prime =(1,-1,-1,1)$ would have an efficiency factor of $\frac{1}{2}$.  In general different treatment contrasts  $\bfe^\prime \hat\bft$ will have different efficiency factors which are given by
\[
\frac{\bfe^\prime {\bf A}\bfe}{\bfe^\prime \bfe},\ \ {\rm where}\\
\]
\begin {equation}
{\bf A}=\frac{1}{r}\tilde\xbp\tilde\xb
=\ib_v-\frac{1}{rk}\nnp.
\label{eq:2.39}
\end {equation}

For the design in Table \ref{table:2.9},

\[
\nnp=
\left[\begin{array}{cccc}
2 & 0 & 1 & 1\\
0 & 2 & 1 & 1 \\
1 & 1 & 2 & 0 \\
1 & 1 & 0 & 2 \\
\end{array}\right]
\label {eq:2.40}
\]
and so
\begin {equation}
{\bf A}=\frac{1}{4}
\left[\begin{array}{rrrr}
2 & 0 & -1 & -1\\
0 & 2 & -1 & -1 \\
-1 & -1 & 2 & 0 \\
-1 & -1 & 0 & 2 \\
\end{array}\right].
\label {eq:2.40}
\end {equation}

Then if $\bfe^\prime =(1,-1,-1,1)$,
\[
\frac{\bfe^\prime {\bf A}\bfe}{\bfe^\prime \bfe}=\frac{1}{2},\\
\]
as expected.

The non-zero latent roots (Definition A.44) of $\bf A$ are known as the canonical efficiency factors ($cef$s) of a design and provide an effective summary of its properties. The $cef$s of the design in Table \ref{table:2.9} are the latent roots of $\bf A$ given by (\ref{eq:2.40}), namely $1, \frac{1}{2}$ and $\frac{1}{2}$ with associated latent vectors $\frac{1}{2}(1,1,-1,-1)^\prime$, $\frac{1}{2}(1,-1,1,-1)^\prime$ and $\frac{1}{2}(1,-1,-1,1)^\prime$ respectively.  A geometrical interpretation of the $cef$s is given by James and Wilkinson (1971).

For a connected design there will be $v-1$ $cef$s $e_1,\ldots, e_{v-1}$. If the associated latent vectors are $\bfe_1,\ldots, \bfe_{v-1}$ then the treatment contrast $\bfe^\prime_i \hat\bft$ will have efficiency factor $e_i$.  The spectral decomposition (Theorem A.33) of ${\bf A}$ can be written as

\begin {equation}
{\bf A}=\sum_{i=0}^{v-1}e_i\mbox{\boldmath $\eta$}_i\mbox{\boldmath $\eta$}_i^\prime,
\label {eq:2.41} 
\end {equation}
where $e_0=0$ and $\bfe_0=\frac{1}{\sqrt v} {\bf 1}_v$.

A unique generalized inverse for $\bf A$, known as the Moore-Penrose inverse (Theorem A.38) is obtained by inverting the $cef$s in (\ref{eq:2.41}).  Hence the variance matrix for $\hat\bft$ can be written in the form

\begin {equation}
{\rm Var}(\hat\bft)=\frac{\sigma^2}{r}\sum_{i=1}^{v-1}\frac{1}{e_i}\mbox{\boldmath $\eta$}_i\mbox{\boldmath $\eta$}_i^\prime .
\label {eq:2.42} 
\end {equation}

The average of the variances of all pairwise contrasts $\hat\bft_i-\hat\bft_{i\prime} (i \ne i^\prime)$ is then given by $\frac{2\sigma^2}{rE}$, where $E$ is called the average efficiency factor and is the harmonic mean of the $cef$s, namely
\begin {equation}
E=\frac{v-1}{\sum_{i=1}^{v-1}\frac{1}{e_i}}.
\label {eq:2.43} 
\end {equation}
The average efficiency factor can be interpreted as the ratio of the variance of a pairwise contrast for two treatments in a randomized complete block design $(\frac {2\sigma^2}{r})$ and the arithmetic average of the variances of all pairwise contrasts for the design.

Since the $e_i$s give the proportion of intra-block information available for the treatment contrast $\bfe_i^\prime \hat\bft$, they are necessarily less than or equal to one.  For an incomplete block design at least one of the $e_i$s will be strictly less than one, so $E$ will also be less than one and provides a useful summary of the amount of information available for the estimation of intr-block treatment means.  Other less commonly used criteria are (i) the geometric mean of the $cef$s and (ii) the smallest $cef$. For a set of basic parameters $v, k$ and $r$, one would normally try to choose a design with the maximum possible value of $E$.  

It is important to note that (i) if blocking is effective in an incomplete block design, then the reduction in the size of the residual mean square compared to that for the comparable randomized complete block design will more than compensate for the loss of information on treatment comparisons (as measured by $E$) and (ii) the intra-block model (\ref{eq:2.34}) can be re-formulated as a mixed model to recover treatment information lost to between-block comparisons.\\

\begin{center}
{\bf 3. Balanced Incomplete Block Designs}
\end{center}
\bigskip
\noindent {\bf 3.1 Description}\\
In a randomized complete block design all possible pairwise comparisons of treatments can be made in each block.  With an incomplete block design this can no longer be done.  However for some incomplete block designs it is still possible for every pairwise treatment comparison to be made within blocks the same number $(\lambda)$ of times.  Such designs are called balanced incomplete block (BIB) designs.  An example of a BIB design is given in Table \ref{table:3.10}.  Each of the  ${7 \choose 2} = 21$ pairwise comparisons between the 7 treatments can be made once within blocks, i.e. the off-diagonal elements of $\nnp$ are all one.

\begin{table}
\centering\small
\caption{{\it A BIB design for 7 treatments in blocks of size 3}}
\vspace{0.2cm}
\begin{tabular}{lccccccc}
{\bf Block} & 1 & 2 & 3 & 4 & 5 & 6 & 7\\
\cline{2-8}
& 1 & 2 & 3 & 4 & 5 & 6 & 7\\
& 2 & 3 & 4 & 5 & 6 & 7 & 1 \\
& 4 & 5 & 6 & 7 & 1 & 2 & 3  \\
\end{tabular}
\label{table:3.10}
\end{table}

In general the concurrence matrix of a BIB design can be written in the form

\begin {equation}
\nnp=(r-\lambda) \ib_v+\lambda\jb_v.
\label {eq:3.44} 
\end {equation}
The row sum of $\nnp$ is thus given as either $rk$ or $r+(v-1)\lambda$. Therefore a condition on the existence of a BIB design is that
\[
\lambda=\frac{r(k-1)}{v-1}
\]
is an integer.

Substitution of (\ref{eq:3.44}) into (\ref{eq:2.39}) gives
\begin {equation}
{\bf A}=\frac{\lambda}{rk}(v\ib_v-\jb_v)=\frac{\lambda v}{rk}\ib_v \ \ ({\rm modulo}\  \jb_v),
\label {eq:3.45} 
\end {equation}
and so the $cef$s for a BIB are all equal to
\begin {equation}
e=\frac{\lambda v}{rk}=\frac{v(k-1)}{k(v-1)}.
\label {eq:3.46} 
\end {equation}
The single $e$ is simply called the efficiency factor for a BIB design and is necessarily less than one; this can be seen by putting $k<v$ in (\ref{eq:3.46}).  Also the number of blocks $b=\frac{vr}{k}$ in a BIB design cannot be less than $v$.  When they exist BIB designs are optimal for all designs with the same basic parameters, $v, k$ and $r$.

The conditions that $\lambda$ is an integer and $b \ge v$ greatly reduce the number of parameter combinations for which the existence of BIB designs is possible.  Furthermore there are a number of combinations of $v, k, r, b$ and $\lambda$ which satisfy the conditions and for which it has been proved that a BIB design does not exist.  For example a BIB design cannot be constructed for $v=b=22, r=k=7$ and $\lambda=2$.

 \bigskip
\noindent {\bf 3.2 Analysis}\\
 For a BIB design the solution of the reduced normal equations (\ref{eq:2.35}) is simple because from (\ref{eq:2.39}), (\ref{eq:3.45}) and (\ref{eq:3.46}),
\begin {equation}
\tilde\xbp\tilde\xb=re\ib_v\ \ ({\rm modulo}\  \jb_v).
\label {eq:3.47} 
\end {equation}
Therefore a generalized inverse for $\tilde\xbp\tilde\xb$ is given by
\begin {equation}
(\tilde\xbp\tilde\xb)^-=\frac{1}{re}\ib_v.
\label {eq:3.48} 
\end {equation}
Hence the vector of adjusted treatment effects for the intra-block analysis of a BIB design is
\begin {equation} 
\hat{\mbox{\boldmath $\tau$}}=\frac{1}{re} \tilde\xb \by,
\label{eq:3.49}
\end {equation}
and the sums of squares for treatments adjusted for blocks is
\begin {equation} 
\byp \tilde\pf_T \by =\frac{1}{re} \byp \tilde\xb\tilde\xbp \by.
\label{eq:3.50}
\end {equation}

From (\ref{eq:3.48}) the variance of a treatment contrast $\bfe^\prime \hat\bft$ is $\frac{\sigma^2}{re} \bfe^\prime \bfe$; for a randomized block design this is $\sigma^2 \bfe^\prime \bfe $ so the variance has been inflated by $\frac{1}{e}$ for a BIB design, reflecting the fact that not all treatments occur in each block.  In particular the variance of any pairwise contrast of estimated means is 
\begin {equation}
 {\rm Var}(\hat{\mbox{\boldmath $\tau$}}_i-\hat{\mbox{\boldmath $\tau$}}_{i\prime})=\frac{2\sigma^2}{re}.
\label{eq:3.51}
\end {equation}

\bigskip
\noindent {\bf 3.3 Factorization of the residual operator}\\
The residual operator (\ref{eq:2.38}) for the linear model (\ref{eq:2.34}) reflects a two-stage procedure for the sweep analysis of any incomplete block design.  Firstly the simple sweep operator $(\ib_n-{\bf P}_B)$ removes block effects ignoring treatments.  Then the usually more complex sweep operator $(\ib_n-\tilde{\bf P}_T)$ removes the effects of treatments adjusted for blocks, leaving just the residual vector $(\ib_n-\tilde{\bf P}_T)(\ib_n-{\bf P}_B) \by$.
This process is represented geometrically in Figure 2; $\sr(\bf Z)$ and $\sr( \bf X)$ are the block and treatment subspaces respectively.  The simple sweep operator  $(\ib_n-{\bf P}_B)$ maps $\by$ into the orthogonal complement of the block subspace within the model subspace; then  $(\ib_n-\tilde{\bf P}_T)$ maps $(\ib_n-{\bf P}_B) \by$ onto the zero vector in the model subspace, i.e. a vector in the residual subspace.

As mentioned in Section 2.2, $\tilde{\bf P}_T$ in general does not have a simple form.  However for a BIB design the residual operator can be further factorized to give a three-stage sweep analysis, i.e.
\beq
&& (\ib_n-\tilde{\bf P}_T)(\ib_n-{\bf P}_B)\\
&=&(\ib_n-\tilde \xb {(\tilde \xbp \tilde \xb)}^- \tilde \xbp)(\ib_n-{\bf P}_B)\\
&=&(\ib_n-\frac{1}{re}\tilde \xb  {\tilde \xbp})(\ib_n-{\bf P}_B) \ \qquad\mbox{(using (\ref{eq:3.47}))}\\
&=&\{\ib_n-\frac{1}{e}(\ib_n-{\bf P}_B){\bf P}_T)(\ib_n-{\bf P}_B)\}(\ib_n-{\bf P}_B) \ \ ({\rm since}\  \tilde \xb=(\ib_n-{\bf P}_B) \xb)\\
\eeq
\begin {equation}
=(\ib_n-{\bf P}_B)(\ib_n-\frac{1}{e}{\bf P}_T)(\ib_n-{\bf P}_B).
\label{eq:3.52}
\end {equation}

\begin{figure}
\centering
\includegraphics [scale=0.5]{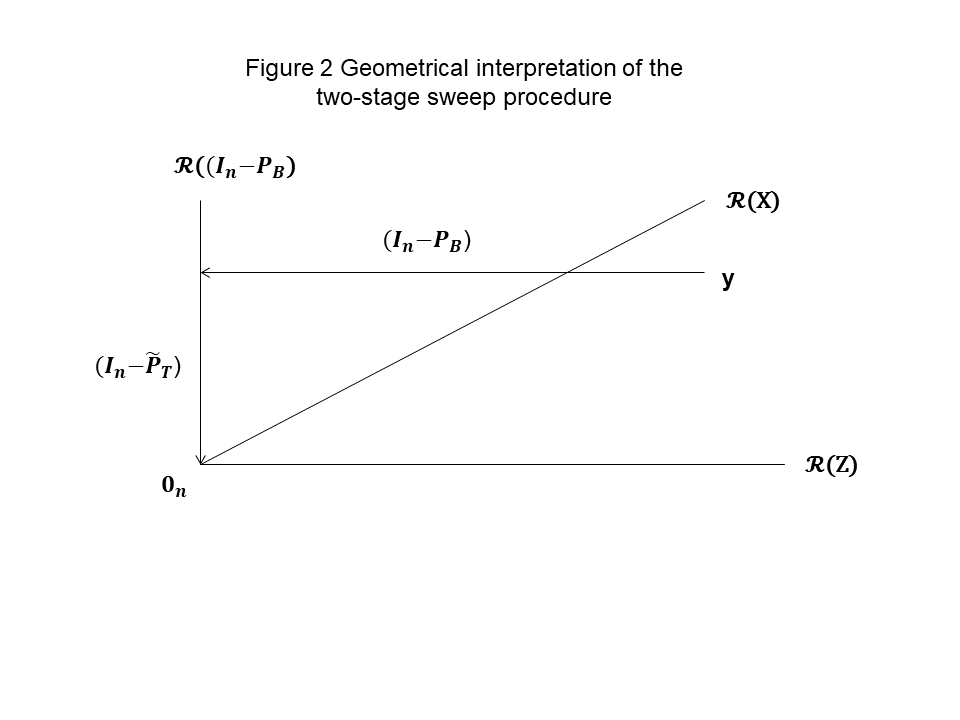}
\label{fig:2}
\end{figure}

Thus for a BIB design the residual operator can be written as a product of a simple sweep for blocks, an augmented sweep for treatments and finally another sweep for blocks.  This procedure is illustrated in Figure 3.

\begin{figure}
\centering
\includegraphics [scale=0.5]{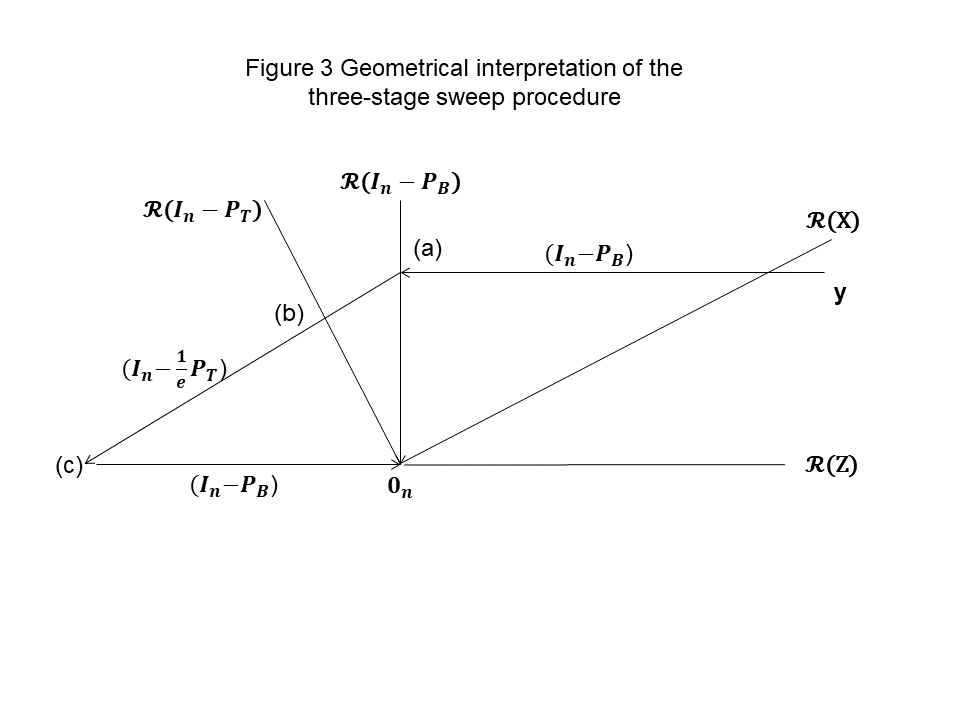}
\label{fig:3}
\end{figure}

As with the two-stage analysis $(\ib_n-{\bf P}_B)$ maps $\yb$ to a point (a) in Figure 3.  The sweep operator $(\ib_n-{\bf P}_T)$ would then map (a) to the point (b) in the orthogonal complement of the treatment subspace within the model subspace.  However the effect of the augmented sweep operator $(\ib_n-\frac{1}{e}{\bf P}_T)$, where the elements of ${\bf P}_T$ have been inflated by the inverse of the efficiency factor, is to extend the mapping to point (c) in the block subspace.  Therefore the final sweep for blocks can map onto the zero vector in the model subspace, as required. 

The decomposition (\ref{eq:3.52}) was first given by James (1957) and forms the cornerstone of the Genstat algorithm for the analysis of non-orthogonal block designs.  Whilst carrying out the sequence of sweeps, the Genstat algorithm also calculates along the way, the vector of adjusted treatment effects
\[
\hat{\mbox{\boldmath $\tau$}}=\frac{1}{e} \xbp {\bf P}_T(\ib_n-{\bf P}_B)\by,
\]
and the sums of squares needed for the analysis of variance table.

\newpage

\begin{center}
\large{\bf \ \ References}
\end{center}
\vspace{0.25cm}

\begin{description}
\item Cochran W.G. (1934).  The distribution of quadratic forms in a normal system with applications to the analysis of covariance.
 {\em Math. Proceedings of the Cambridge Philosophical Society} , 178--191.

\item James A.T. (1957). The relationship algebra of an experimental design.
{\em Ann. Math. Statist.} {\bf 28}, 993--1002.

\item James A.T. \& Wilkinson G.N. (1971). Factorization of the residual operator and canonical decomposition of nonorthogonal factors in analysis of variance.
{\em Biometrika} {\bf 58}, 279--294.

\item Madow W.G. (1940).  Limiting distributions of quadratic and bilinear forms.
{\em Ann. Math. Statist.} {\bf 11}, 125--146.

\item Nelder J.A. (1977). A reformulation of linear models. {\em J. Roy. Statist. Soc. A.} {\bf 140}, 48--63.

\item Searle S.R. (1971). {\em Linear Models.} New York: Wiley.

\end{description}

\newpage

\begin{center}
\large{\bf \ \ Appendix A. Vector Spaces and Matrices}
\end{center}
\vspace{0.25cm}

\noindent {\bf A.1 Vector spaces}\\
\noindent{\it Definition A.1.} A real vector space $V$ is a set of vectors for 
which the operations of addition ($+$) and scalar multiplication by a real 
number satisfy the following conditions:
\medskip
\begin{enumerate}
\item[(i)] If $\bv_1,\bv_2\in V$ then $\bv_1+\bv_2$ is a unique vector $\in V$.
\item[(ii)] $(\bv_1+\bv_2)+\bv_3=\bv_1+(\bv_2+\bv_3)$ for all 
$\bv_1,\bv_2,\bv_3\in V$.
\item[(iii)] $\bv_1+\bv_2=\bv_2+\bv_1$ for all $\bv_1,\bv_2\in V$. 
\item[(iv)] There is a zero vector $\ze\in V$ such that $\bv+\ze=\ze+\bv=\bv$ 
for all $\bv\in V$.
\item[(v)] If $\bv\in V$ there exists a vector $-\bv\in V$ such that $\bv+(-
\bv)=(-\bv)+\bv=\ze$.
\item[(vi)] If $\bv\in V$ and $a$ is any real number, i.e.\ $a\in\sr$, then 
$a\bv$ is a unique vector $\in V$.
\item[(vii)] $a(b\bv)=b(a\bv)=(ab)\bv$ for all $a,b\in\sr$ and $\bv\in V$.
\item[(viii)] $(a+b)\bv=a\bv+b\bv$ for all $a,b\in\sr$ and $\bv\in V$.
\item[(ix)] $a(\bv_1+\bv_2)=a\bv_1+a\bv_2$ for all $a\in\sr$ and 
$\bv_1,\bv_2\in V$.
\item[(x)] $1\bv=\bv$ for all $\bv\in V$.
\end{enumerate}
\medskip
Henceforth  a real vector space will be referred to simply as a vector space.

\medskip
\noindent{\it Definition A.2.} A subspace $U$ of a vector space $V$ is a set of 
vectors of $V$ which satisfy the following conditions:
\begin{enumerate}
\item[(i)] If $\bu_1,\bu_2\in U$ then $\bu_1+\bu_2\in U$.
\item[(ii)] If $\bu\in U$ then $a\bu\in U$ for all $a\in\sr$.
\item[(iii)] If $\ze\in V$ then $\ze\in U$.
\end{enumerate}

\medskip
\noindent{\it Definition A.3.} The sum $U_1+U_2$ of two subspaces $U_1$, $U_2$ 
of a vector space $V$ is the set 
\[U_1+U_2=\{\bv=\bu_1+\bu_2\mid\bu_1\in U_1\;,\;\bu_1\in U_2\}\;.\]

\medskip
\noindent{\it Definition A.4.} The intersection $U_1\cap U_2$ of two subspaces 
$U_1,U_2$ of a vector space $V$ is the set \[U_1\cap U_2=\{\bu\mid\bu\in 
U_1\;,\;\bu\in U_2\}\;.\]

The sum and the intersection of two subspaces of $V$ are themselves subspaces 
of $V$. 

\medskip
\noindent{\it Definition A.5.} Two subspaces are called disjoint if their intersection 
is only the zero vector, i.e. \[U_1\cap U_2=\ze\;.\]

\medskip
\noindent{\it Definition A.6.} If $U$ is a subspace of a vector space $V$ then 
for any $\bv\in V$, the subset \[\bv+U=\{\bv+\bu\mid\bu\in U\}\] is called a 
coset of $U$ in $V$.

\noindent{\it Theorem A.1.} If $\bv_2$ is any other vector in the coset $\bv_1+U$ 
then $\bv_1+U=\bv_2+U$.

\noindent{\it Proof.} If $\bv_2\in\bv_1+U$ then there exists a vector $\bu_1\in 
U$ such that $\bv_2=\bv_1+\bu_1$. Suppose that $\bv_3\in\bv_2+U$; then there 
exists a vector $\bu_2\in U$ such that 
\begin{eqnarray*}
\bv_3&=&\bv_2+\bu_2\\
&=&\bv_1+\bu_2+\bu_2\\
&=&\bv_1+\bu_3\;,\quad\mbox{where $\bu_3\in U$}\;.
\end{eqnarray*}
Hence $\bv_2+U\subset\bv_1+U$. 
By a similar argument $\bv_1+U\subset\bv_2+U$ and hence $\bv_1+U=\bv_2+U$. 

The subspace $U$ can be interpreted as a hyperplane passing through the zero 
vector $\ze$ or origin of $V$. The coset $\bv_1+U$ is a translation of $U$ 
which can be interpreted as a hyperplane through $\bv_1$ parallel to $U$. This 
is illustrated in Figure A1.

\begin{figure}
	\centering
		\includegraphics [scale=0.5]{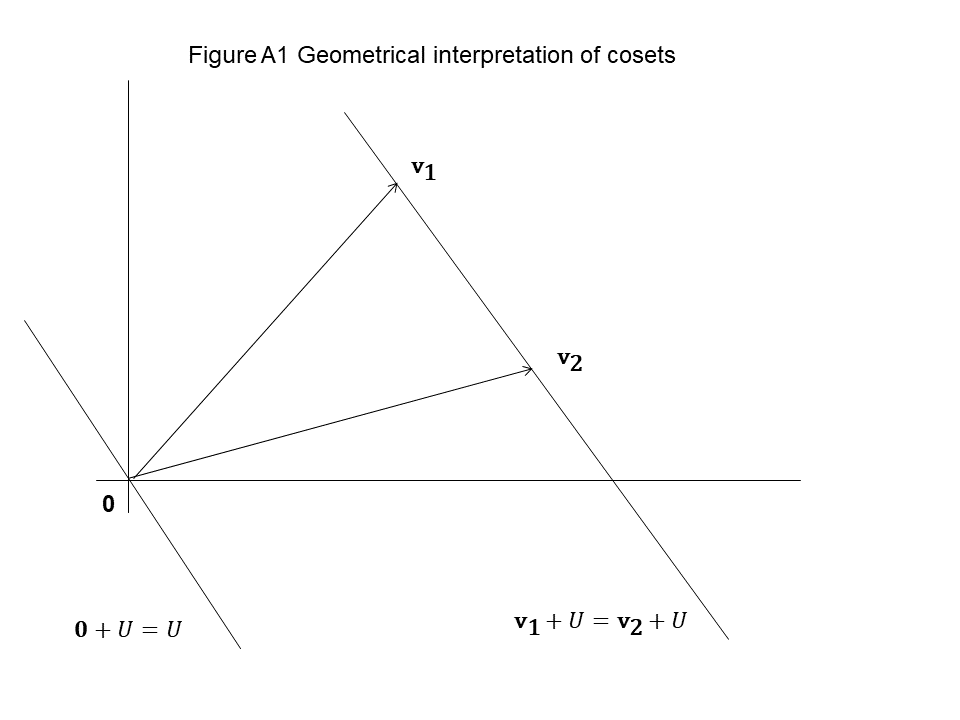}
	\label{fig:A1}
\end{figure}

The hyperplanes passing through $\bv_1$ and $\bv_2$ parallel to $U$ are the 
same.

\noindent{\it Definition A.7.} If $U$ is a subspace of a vector space $V$ then 
the set 
\[\{\bv+U\mid\bv\in V\}\]
of cosets is called the quotient space, $V\mid U$.

\noindent{\it Theorem A.2.} The quotient space $V\mid U$ forms a vector space 
under the operation of addition defined by  
\[(\bv_1+U)+(\bv_2+U)=(\bv_1+\bv_2)+U\;,\]
and scalar multiplication defined by 
\[a(\bv+U)=a\bv+U\;,\quad a\in\sr\;.\]

\noindent{\it Definition A.8.} A cross-section of the quotient space $V\mid U_1$ 
of a subspace $U_1$ in a vector space $V$ is a subspace $U_2$ of $V$, disjoint 
from $U_1$ such that $V$ is the sum of $U_1$ and $U_2$, i.e.\ 
\[V=U_1+U_2\;,\quad U_1\cap U_2=\ze\;.\]

\bigskip
\noindent{\bf A.2 Linear transformations}\\
\noindent{\it Definition A.9.} A linear transformation $T$ from a vector space 
$V$ to a vector space $W$ is a mapping 
\[\bv\in V\buildrel\rm T\over \to \bw=T(\bv)\in W\;,\]
from $V$ into $W$ such that:
\begin{enumerate}
\item[(i)] $T(\bv_1+\bv_2)=T(\bv_1)+T(\bv_2)\;,\quad \bv_1,\bv_2\in V$
\item[(ii)] $T(a\bv)=aT(\bv)\;,\quad a\in\sr\;,\;\bv\in V\;.$
\end{enumerate}

\noindent{\it Definition A.10.} The range $\sr(T)$ of a linear transformation 
$T$ from $V$ to $W$ is the set of vectors $\bw\in W$ which can be written in 
the form $\bw=T(\bv)$ for some $\bv\in V$.

The range of $T$ is a subspace of $W$.

\noindent{\it Definition A.11.} The kernel $\ck(T)$ or null space of a linear 
transformation $T$ from $V$ to $W$ is the set of vectors $\bv\in V$ which are 
mapped to the zero vector of $W$, i.e. 
\[\ck(T)=\{\bv\in V\mid T(\bv)=\ze\}\;.\]
The kernel of $T$ is a subspace of $V$.

\bigskip
\noindent{\bf A.3 Generalized inverses}\\
\noindent{\it Definition A.12.}  A generalized inverse of a linear 
transformation $T$ from a vector space $V$ to a vector space $W$ is a linear 
transformation $S$ from $W$ to $V$ such that for any arbitrary vector $\bw\in 
\sr(T)\subset W$, the vector $\bv=S(\bw)\in V$ is a solution of the equation 
$T(\bv)=\bw$.

\noindent{\it Theorem A.3.} The linear transformation $S$ is a generalized inverse 
of $T$ if and only if $TST=T$. 

\noindent{\it Proof.} By definition $\bv=S(\bw)$ is a solution of $T(\bv)=\bw$ 
for all $\bw\in\sr(T)$
\begin{eqnarray*}
&\llra&TS(\bw)=\bw\qquad\mbox{for all $\bw\in\sr(T)$}\\
&\llra&TST(\bv)=T(\bv)\qquad\mbox{for all $\bv\in V$}\\
&\llra&TST=T\;.
\end{eqnarray*}

\noindent{\it Theorem A.4.} A generalized inverse $S$ of a linear transformation 
$T$ from a vector space $V$ to a vector space $W$ is a linear transformation 
from $W$ to $V$.

A geometrical interpretation of a generalized inverse is given in Figure A2.

\begin{figure}
	\centering
		\includegraphics [scale=0.6]{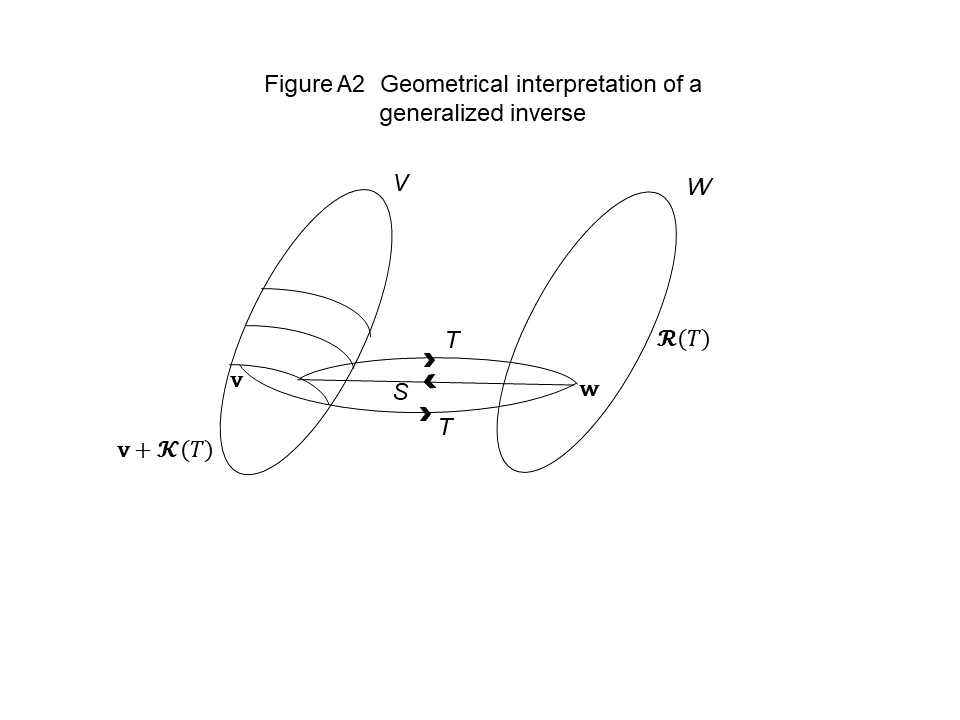}
	\label{fig:A2}
\end{figure}

When $W$ is restricted to $\sr(T)$ the generalized inverse $S$ produces a one 
to one mapping of $\sr(T)$ onto some cross-section $U$ of $V\mid\ck(T)$.

\bigskip
\noindent{\bf A.4 Projections}\\

\noindent{\it Theorem A.5.} If a vector space $V$ is the sum of two disjoint subspaces 
$U_1$ and $U_2$ then any vector $\bv=\bu_1+\bu_2\in V$ uniquely determines the 
vectors $\bu_1\in U_1$ and $\bu_2\in U_2$. 

\noindent{\it Proof.} Suppose $\bv$ can be written as 
\[\bv=\bu_1^*+\bu_2^*\;,\quad\bu_1^*\in U_1\;,\;\bu_2^*\in U_2\;.\]
Then on subtraction 
\[\ze=(\bu_1-\bu_1^*)+(\bu_2-\bu_2^*)\;.\]
Thus $(\bu_1-\bu_1^*)=-(\bu_2-\bu_2^*)\in U_1\cap U_2$, and hence $(\bu_1-
\bu_1^*)=\ze=-(\bu_2-\bu_2^*)$, i.e.\ $\bu_1=\bu_1^*$ and $\bu_2=\bu_2^*$.

\noindent{\it Definition A.13.} Let the vector space $V$ be the sum of two 
disjoint subspaces $U_1$ and $U_2$, i.e.\ $\bv=\bu_1+\bu_2$, \ $\bv\in V$,
$\bu_1\in U_1$, $\bu_2\in U_2$. The mapping $P$ from $V$ to $U_1$ defined by 
\[\bv\in V\buildrel{\rm P}\over \to\bu_1=P(\bv)\in U_1\]
is called the projection of $V$ on $U_1$ parallel to $U_2$. 

A projection is a linear transformation.

\noindent{\it Theorem A.6.} A linear transformation $P$ of a vector space $V$ into itself 
is a projection if and only if it is idempotent, i.e.\ $P^2=P$. 

\noindent{\it Proof.} (i) Supose $P$ is a projection on $U_1$ parallel to 
$U_2$, where $V=U_1+U_2$, $U_1\cap U_2=\ze$ and $U_1,U_2\subset V$. If 
$\bv=\bu_1+\bu_2$, \ $\bv\in V$, $\bu_1\in U_1$, $\bu_2\in U_2$, then 
$P(\bv)=\bu_1$ and thus $P(\bu_1)=\bu_1$. Therefore 
\[P^2(\bv)=PP(\bv)=P(\bu_1)=\bu_1=P(\bv)\;,\]
for all $\bv\in V$. Hence $P^2=P$. 

(ii) Now let $P$ be idempotent and set $U_1=\sr(P)$ and $U_2=\sr(I-P)$, where 
$I$ is the identity transformation, i.e.\ $I(\bv)=\bv$. 
For any arbitrary vector $\bv\in V$, 
\[\bv=P(\bv)+(I-P)(\bv)\;.\]
Hence $V=U_1+U_2$. It remains to show that $U_1$ and $U_2$ are disjoint 
subspaces. Suppose $\bv\in U_1\cap U_2$, then there exist vectors $\bv_1$ and 
$\bv_2$ such that $\bv=P(\bv_1)=(I-P)(\bv_2)$. Therefore 
\[\bv=P(\bv_1)=P^2(\bv_1)=P(\bv)=P(I-P)(\bv_2)=\ze\;.\]
Hence $P$ is a projection on $U_1$ parallel to $U_2$. As a result the following 
theorem can then be stated:

\noindent{\it Theorem A.7.} If $P$ is a projection of a vector space $V$, then
\begin{enumerate}
\item[(i)] $\sr(I-P)=\ck(P)$
\item[(ii)] $\ck(I-P)=\sr(P)$
\item[(iii)] $\sr(P)\cap\ck(P)=\ze$
\item[(iv)] $\sr(P)+\ck(P)=V$.
\end{enumerate}

\bigskip
\noindent{\bf A.5 Finite dimensional vector spaces}\\
\noindent{\it Definition A.14.} A set of vectors $\bv_1,\ldots,\bv_n$ of a 
vector space $V$ is a linearly dependent set if there exist scalars 
$a_1,\ldots,a_n \in\sr$, not all zero such that 
\[a_1\bv_1+\cdots+a_n\bv_n=\ze\;.\]
A set of vectors which are not linearly dependent are said to be linearly 
independent.

\noindent{\it Theorem A.8.} Let $\bv_1,\ldots,\bv_n$ be vectors of a vector space $V$. 
The set of all linear combinations of $\bv_1,\ldots,\bv_n$ is a subspace $U$ of 
$V$, i.e. 
\[U=\{\bu=a_1\bv_1+\cdots+a_n\bv_n\mid a_1,\ldots,a_n\in\sr\}\;.\]
The subspace $U$ is said to be generated by the vectors $\bv_1,\ldots,\bv_n$.

\noindent{\it Definition A.15.} Let $U$ be the subspace of a vector space $V$ 
generated by the vectors $\bv_1,\ldots,\bv_n\in V$. The set 
$\bv_1,\ldots,\bv_n$ is called a basis of $U$ if it is a linearly independent 
set.

It can be shown that any two bases of a vector space have the same number of 
vectors.

\noindent{\it Definition A.16.} The number of vectors in a basis of a vector 
space $V$ is called the dimension of $V$ and is denoted by ${\rm dim}\,(V)$. 

\bigskip
\noindent{\bf A.6 Matrices}\\
\noindent{\it Definition A.17.} An $n\times p$ matrix $\hb$ is a rectangular 
array of real numbers $h_{ij}$ \ $(i=1,\ldots,n\,;\,j=1,\ldots,p)$. 

Only real matrices will be considered here and henceforth they will be 
referred to simply as matrices. An $n\times p$ matrix is said to have $n$ rows 
and $p$ columns. The columns of an $n\times p$ matrix can be thought of as a 
set of $p$ vectors of length $n$. 

\noindent{\it Definition A.18.} The transpose $\hb^\prime$ of an $n\times p$ 
matrix $\hb$ is a $p\times n$ matrix whose $(j,i)$th element is equal to 
the $(i,j)$th element of $\hb$ \ $(i=1,\ldots,n\,;\,j=1,\ldots,p)$.

\noindent{\it Definition A.19.} An $n\times n$ matrix is called a square matrix 
of order $n$. 

\noindent{\it Definition A.20.} A symmetric matrix is a square matrix of order 
$n$ whose elements $h_{ij}$ satisfy the relation $h_{ij}=h_{ji}$ \ 
$(i,j=1,\ldots,n)$.

\noindent{\it Definition A.21.} Let $\gb$ and $\hb$ be $n\times p$ matrices 
with elements $g_{ij}$ and $h_{ij}$ respectively 
$(i=1,\ldots,n\,;\,j=1,\ldots,p)$. The sum $\gb+\hb$ of $\gb$ and $\hb$ is an 
$n\times p$ matrix whose $(i,j)$th element is $g_{ij}+h_{ij}$.

\noindent{\it Definition A.22.} Let $\gb$ be a $q\times n$ matrix and $\hb$ an 
$n\times p$ matrix with elements $g_{ki}$ and $h_{ij}$  respectively  
$(i=1,\ldots,n\,;\,j=1,\ldots,p\,;\,k=1,\ldots,q)$. The product $\gb\hb$ of 
$\gb$ and $\hb$ is a $q\times p$ matrix whose $(k,j)$th element is 
$\sum_{i=1}^n g_{ki}h_{ij}$.

\noindent{\it Definition A.23.} A diagonal matrix of order $n$ is a square 
matrix will all off-diagonal elements equal to zero and is written as 
diag$(a_1,\ldots,a_n)$, where $a_1,\ldots,a_n$ are the diagonal elements.

\noindent{\it Definition A.24.} A diagonal matrix of order $n$ with all 
diagonal elements equal to one is calleld the identity matrix of order $n$ and 
is written as $\ib_n$.
It is also useful to identify two other types of matrix:
\begin{enumerate}
\item[(i)] An $n\times p$ matrix with all elements equal to zero is written as 
$\ze_{n,p}$.
\item[(ii)] An $n\times p$ matrix with all elements equal to one is written as 
$\jb_{n,p}$, however for a square matrix of order $n$, this is abbreviated to 
simply $\jb_n$.
\end{enumerate}

\noindent{\it Definition A.25.} The trace of a square matrix $\hb$ of order $n$ 
is the sum of the diagonal elements of $\hb$, i.e.\ trace$(\hb)=\sum_{i=1}^n 
h_{ii}$.
Some properties of the trace of a matrix are summarized in the following two 
theorems:

\noindent{\it Theorem A.9.} Let $\gb$ and $\hb$ the square matrices of order $n$. Then 
\begin{enumerate}
\item[(i)] trace$(\hb^\prime)=\mbox{trace}(\hb)$,
\item[(ii)] trace$(\gb+\hb)=\mbox{trace}(\gb)+\mbox{trace}(\hb)$, 
\item[(iii)] trace$(a\hb)=a\mbox{trace}\,(\hb)$ for $a\in \sr$.
\end{enumerate}

\noindent{\it Theorem A.10.} Let $\gb$ be a $p\times n$ matrix and $\hb$ an $n\times 
p$ matrix. Then 
\[\mbox{trace}\,(\gb\hb)=\mbox{trace}\,(\hb\gb)\;.\]

\noindent{\it Definition A.26} The column (row) rank of a matrix $\hb$ is the 
number of linearly independent columns (rows) of $\hb$.

\noindent{\it Theorem A.11.} The column rank and row rank of a matrix $\hb$ are equal.

Hence the column (row) rank of $\hb$ can simply be called the rank of $\hb$, 
denoted by rank $(\hb)$. 

\noindent{\it Definition A.27.} An independent matrix $\hb$ is a square matrix 
with the property that \[\hb^2=\hb\;.\]

\noindent{\it Definition A.28.} An orthogonal matrix $\hb$ is a square matrix 
of order $n$ with the property that 
\[\hb^\prime\hb=\ib_n\;.\]

\bigskip
\noindent{\bf A.7 Linear transformations as matrices}\\
The preceding theory has been for vector spaces in general. However for the 
purpose of this Appendix one need only consider the finite dimensional vector space 
$\sr^n$ of vectors of length $n$, i.e.\ $\bx^\prime=(x_1,\ldots,x_n)$, \ 
$x_i\in\sr$ $(i=1,\ldots,n)$. Then a linear transformation from, say $\sr^p$ to 
$\sr^n$ can be written as an $n\times p$ matrix $\tb$ which maps a vector 
$\bx\in\sr^p$ to a vector $\by\in\sr^n$, i.e.\ 
\[\by=\tb\bx\;.\]

\noindent{\it Theorem A.12.} If a linear transformation from $\sr^p$ to $\sr^n$ is 
written as an $n\times p$ matrix $\tb$ then the range of $\tb$, $\sr(\tb)$, is 
the subspace generated by the columns of $\tb$. 

\noindent{\it Theorem A.13.} If a linear transformation is 
written as a matrix $\tb$ then the dimension of $\sr(\tb)$, dim\,$(\sr(\tb))$ 
is equal to rank $(\tb)$.

\noindent{\it Theorem A.14.} If a linear transformation from $\sr^p$ to $\sr^n$ is 
written as an $n\times p$ matrix $\tb$ then 
dim\,$(\sr(\tb))+\mbox{dim}\,(\ck(\tb))=n$.

\noindent{\it Theorem A.15.} Suppose a linear transformation from $\sr^p$ to $\sr^n$ is 
written as an $n\times p$ matrix $\tb$. If ${\bf S}$ is the $p\times n$ matrix of a 
generalized inverse transformation, then $\tb{\bf S}\tb=\tb$.

The generalized inverse of a matrix $\tb$ is usually denoted as $\tb^-$. 

\bigskip
\noindent{\bf A.8 Orthogonal subspaces}\\
\noindent{\it Definition A.29.} The scalar product $\bv_1^\prime\bv_2$ of two 
vectors $\bv_1,\bv_2\in\sr^n$ is defined as 
\[\bv_1^\prime\bv_2=\sum_{i=1}^n\bv_{1i}\bv_{2i}\;,\]
where $\bv_{j1},\ldots,\bv_{jn}$ are the elements of $\bv_j$ \ $(j=1,2)$.

\noindent{\it Theorem A.16.} For a vector $\bv\in\sr^n$, \ $\bv^\prime\bv\ge 0$ with 
equality if and only if $\bv=\ze_n$. 

\noindent{\it Definition A.30.} Two vectors $\bv_1,\bv_2\in\sr^n$ are 
orthogonal if their scalar product is zero, i.e.\ $\bv_1^\prime\bv_2=0$.

\noindent{\it Definition A.31.} A vector $\bv\in\sr^n$ is said to be normalized 
if $\bv^\prime\bv=1$. Thus any vector $\bv\in\sr^n$ can be normalized by 
dividing by the scalar $\sqrt{\bv^\prime\bv}$. 

\noindent{\it Definition A.32.} A set of vectors $\bv_1,\ldots,\bv_p\in\sr^n$ 
form an orthogonal set if the vectors are \ (i) normalized and \ (ii) pairwise 
orthogonal. 

\noindent{\it Theorem A.17.} The rows (columns) of an orthogonal matrix of order $n$ 
form an orthogonal basis for $\sr^n$.

\noindent{\it Definition A.33.} A vector $\bv\in\sr^n$ is orthogonal to a 
subspace $U\subset\sr^n$ if it is orthogonal to every vector in the subspace.

\noindent{\it Definition A.34.} Two subspaces $U_1,U_2\subset\sr^n$ are 
orthogonal, $U_1\perp U_2$, if every vector in $U_1$ is orthogonal to every 
vector in $U_2$. 

\noindent{\it Definition A.35.} The orthogonal complement $U^{\perp}$ of a 
subspace $U\subset\sr^n$ is the set of all vectors orthogonal to $U$. 

The orthogonal complement of a subspace is also a subspace.

\noindent{\it Theorem A.18.} If $\tb$ is the matrix of a linear transformation from 
$\sr^n$ to a subspace $U\subset\sr^n$, then $U^{\perp}=\ck(\tb^\prime)$.

\noindent{\it Proof.} From Definition A.11, $\ck(\tb^\prime)$ is the set of all 
vectors $\bv\in\sr^n$ such that $\tb^\prime\bv=\ze$, i.e.\ 
$\bv^\prime\tb=\ze^\prime$. Thus if $\bv\in\ck(\tb^\prime)$ then $\bv$ is 
orthogonal to every column of $\tb$ and hence orthogonal to $U=\sr(\tb)$. So 
from Definition A.35, $U^{\perp}=\ck(\tb^\prime)$. 

The following theorem can then be stated:

\noindent{\it Theorem A.19.} If $\tb$ is a symmetric matrix then 
\[\sr(\tb)^{\perp}=\ck(\tb)\;.\]

\noindent{\it Definition A.36.} The sum of two orthogonal subspaces 
$U_1,U_2\subset\sr^n$ is called the direct sum of $U_1$ and $U_2$ and is denoted 
by $U_1\oplus U_2$. 

\noindent{\it Definition A.37.} If $U_1$ is a subspace of $U_2$ which is a 
subspace of $\sr^n$, i.e.\ $U_1\subset U_2\subset \sr^n$, then the set of all 
vectors of $U_2$ orthogonal to $U_1$ is called the orthogonal complement of 
$U_1$ in $U_2$ and is denoted by $U_2\ominus U_1$. 
The orthogonal complement of $U_1$ in $U_2$ is a subspace of dimension 
$\mbox{dim}\,(U_2)-{\rm dim}\,(U_1)$. 

\noindent{\it Theorem A.20.} If $U$ is a subspace of $\sr^n$ then 
\[\sr^n=U\oplus U^{\perp}\;.\]

\noindent{\it Proof.} If ${\rm dim}\,(U)=p$ then from Theorems A.14 and A.18, ${\rm dim}\,(U^{\perp})=n-p$. Hence $U\oplus U^{{\perp}}$ is a 
subspace of $\sr^n$ of dimension $p+(n-p)=n$ and so is equal to $\sr^n$. 

\bigskip
\noindent{\bf A.9 Orthogonal projections}\\
\noindent{\it Definition A.38.} A projection of $\sr^n$ with matrix $\pf$ is 
orthogonal if $\sr(\pf)\perp\ck(\pf)$.

\noindent{\it Theorem A.21.} The matrix $\pf$ of a projection of $\sr^n$ is the matrix 
of an orthogonal projection if and only if it is symmetric.

\noindent{\it Proof.} (i) If $\pf$ is symmetric then from Theorem A.19 it is 
the matrix of an orthogonal projection.

(ii) Now suppose that $\pf$ is the matrix of an orthogonal projection. It is 
sufficient to prove that the scalars $\bv_1^\prime\pf\bv_2$ and 
$\bv_2^\prime\pf\bv_1$ are equal for arbitrary $\bv_1,\bv_2\in\sr^n$, because 
then $\bv_1^\prime\pf\bv_2=\bv_1^\prime\pf^\prime\bv_2$.

\begin{eqnarray*}
{\rm Let}\qquad \bv_i
&=&\pf\bv_i+(\ib_n-\pf)\bv_i\\
&=&\bv_{i1}+\bv_{i2}\;,\qquad{\rm say} \\
{\rm where}\qquad\bv_{i1}
&\in&\sr(\pf)\quad{\rm and}\quad\bv_{i2}\in\ck(\pf)\;,\quad(i=1,2)\;.
\end{eqnarray*}
Then since $\bv_{i1}^\prime\bv_{j2}=0$ for $i,j=1,2$ and $\pf(\ib_n-
\pf)=\ze_{n,n}$, 
\begin{eqnarray*}
\bv_1^\prime\pf\bv_2
&=&\bv_{11}^\prime\bv_{21}\\
&=&\bv_{21}^\prime\bv_{11}=\bv_2^\prime\pf\bv_1\;.
\end{eqnarray*}
\[{\rm Hence}\qquad\pf^\prime=\pf\;.\]
The preceding theory is summarized in the following theorem:

\noindent{\it Theorem A.22.} A square matrix $\pf$ of order $n$ produces an orthogonal 
projection on a subspace $U\subset\sr^n$ if and only if $\pf$ is a symmetric 
idempotent matrix whose range is $U$.
 
\noindent{\it Theorem A.23.} Two subspaces of $\sr^n$ are orthogonal if and only if the 
product of the matrices $\pf,\qb$ of the orthogonal projections on them is 
zero, i.e.\ 
\[\sr(\pf)\perp\sr(\qb)\llra\pf\qb=\qb\pf=\ze_{n,n}\;.\]

\noindent{\it Proof.} $\sr(\pf)\perp\sr(\qb)$
\begin{eqnarray*}
&\llra&\sr(\qb)\subset\sr(\pf)^{{\perp}}=\ck(\pf)\\
&\llra&\pf\qb=\ze_{n,n}\\
&\llra&\qb\pf=\ze_{n,n}\qquad\mbox{(since $\ze_{n,n}$ is symmetric)}\;.
\end{eqnarray*}

The following result can then be obtained from the previous two theorems. 

\noindent{\it Theorem A.24.} The direct sum of two subspaces of $\sr^n$ with orthogonal 
projection matrices $\pf$ and $\qb$ respectively, has orthogonal, projection 
matrix $\pf+\qb$.
 
\noindent{\it Theorem A.25.} Let $\pf$ and $\qb$ be idempotent symmetric matrices of 
order $n$ whose ranges intersect in the subspace $U=\sr(\pf)\cap\sr(\qb)$. Then 
the orthogonal complements $\sr(\pf)\ominus U$ and $\sr(\qb)\ominus U$ of $U$ 
in $\sr(\pf)$ and $\sr(\qb)$ respectively are orthogonal subspaces of $\sr^n$ 
if and only if $\pf$ and $\qb$ commute, i.e.\ 
\[\pf\qb=\qb\pf\;.\]

\noindent{\it Proof.}
Let $\pf_1,\qb_1$ and $\mb$ be the matrices of the orthogonal projections on 
$\sr(\pf)\ominus U$, $\sr(\qb)\ominus U$ and $U$ respectively. From Theorem 
A.24 it follows that $\pf=\pf_1+\mb$ and $\qb=\qb_1+\mb$. Also from Theorem A.23, 
$\pf_1\mb=\mb\pf_1=\ze=\qb_1\mb=\mb\qb_1$. Hence
\begin{eqnarray*}
\sr(\pf_1)
&=&\sr(\pf)\ominus U\perp \sr(\qb_1)=\sr(\qb)\ominus U\\
&\llra&\pf_1\qb_1=\qb_1\pf_1=\ze_{n,n}\\
&\llra&(\pf_1+\mb)(\qb_1+\mb)=(\qb_1+\mb)(\pf_1+\mb)\\
&\llra&\pf\qb=\qb\pf\;.
\end{eqnarray*}

\bigskip
\noindent{\bf A.10 Determinants}\\
The determinant of a square matrix of order $n$ is most easily defined in terms 
of the determinants of a linear combination of determinants of submatrices of 
order $n-1$. Then in addition only the definition of, say, the determinant of a 
$2\times 2$ matrix is needed. 

\noindent{\it Definition A.39.} The determinant of a $2\times 2$ matrix 
\[\hb=\left(\begin{array}{cc}
h_{11}&h_{12}\\
h_{21}&h_{22}
\end{array}\right)\]
is given by 
\[{\rm det}\,(\hb)=h_{11}h_{22}-h_{12}h_{21}\;.\]

\noindent{\it Definition A.40.} Let $\hb$ be a square matrix of order $n$ with 
elements $h_{ij}$ $(i,j=1,\ldots,n)$. The cofactor $\hb_{ij}$ of an element 
$h_{ij}$ is defined as the determinant of the submatrix of order $n-1$ obtained 
by deleting the $i$th row and $j$th column of $\hb$, multiplied by $(-
1)^{i+j}$.

\noindent{\it Definition A.41.} The determinant of a square matrix $\hb$ of 
order $n$ with elements $h_{ij}$ $(i,j=1,\ldots,n)$ is given by 
\[\det(\hb)=\sum_{i=1}^n h_{ij}\hb_{ij}\;,\quad\mbox{for any $j$}\;.\]

Alternatively by concentrating on the $i$th row of $\hb$, its determinant can 
be written as 
\[\det(\hb)=\sum_{j=1}^n h_{ij}\hb_{ij}\;.\]

\noindent{\it Theorem A.26.} If $\gb$ and $\hb$ are square matrices of order $n$, then 
$\det(\gb\hb)=\det(\hb\gb)=\det(\gb)\det(\hb)$. 

\noindent{\it Definition A.42.} If for a square matrix $\hb$, $\det(\hb)=0$ 
then $\hb$ is said to be singular; if $\det(\hb)\ne 0$ then $\hb$ is non-%
singular. 

\noindent{\it Definition A.43.} The inverse $\hb^{-1}$ of a non-singular matrix 
$\hb$ of order $n$ is a square matrix of order $n$ which satisfies 
\[\hb^{-1}\hb=\hb\hb^{-1}=\ib_n\;.\]

\noindent{\it Theorem A.27.} The $(i,j)$th element of the inverse of a non-singular 
matrix $\hb$ of order $n$ is given by 
\[\frac{1}{\det(\hb)}\hb_{ji}\;,\quad(i,j=1,\ldots,n)\;.\]

\bigskip
\noindent{\bf A.11 Latent roots and latent vectors}\\
\noindent{\it Definition A.44.} Let $\hb$ be a symmetric matrix of order $n$. 
If $\lambda$ is a scalar and $\mbox{\boldmath $\eta$}$ is a vector $\in\sr^n$ such that 
$\hb\mbox{\boldmath $\eta$}=\lambda\mbox{\boldmath $\eta$}$, then $\lambda$ is called a latent root of $\hb$ with 
associated latent vector $\mbox{\boldmath $\eta$}$. 

Only the latent roots of symmetric matrices are considered here. 
It can then be shown that the latent roots are real numbers. 

\noindent{\it Theorem A.28.} The set of all latent roots of a symmetric matrix $\hb$ of 
order $n$ are the solutions of the determinantal equation
\[\det(\ib_n-\lambda\hb)=0\;.\]

\noindent{\it Definition A.45.} A symmetric matrix $\hb$ of order $n$ with 
latent roots $\lambda_i$ $(i=1,\ldots,n)$ is called positive definite if all 
the $\lambda_i$ are greater than zero; if all the $\lambda_i$ are non-negative 
$\hb$ is said to be positive semi-definite.

\noindent{\it Theorem A.29.} If $\hb$ is an $n\times p$ matrix then $\hb\hb^\prime$ and 
$\hb^\prime\hb$ are positive semi-definite matrices of order $n$ and $p$ 
respectively.

\noindent{\it Theorem A.30.} If $H$ is an $n\times p$ matrix then the non-zero latent 
roots of $\hb\hb^\prime$ and $\hb^\prime\hb$ are the same. 

\noindent{\it Proof.} Let $\lambda$ be a latent root of $\hb\hb^\prime$ with 
associated latent vector $\mbox{\boldmath $\eta$}$. Then 
\[\hb\hb^\prime\mbox{\boldmath $\eta$}=\lambda\mbox{\boldmath $\eta$}\;,\]
\[{\rm thus}\qquad\hb^\prime\hb(\hb^\prime\mbox{\boldmath $\eta$})=\lambda(\hb^\prime\mbox{\boldmath $\eta$})\;.\]
and hence $\lambda$ is also a latent root of $\hb^\prime\hb$ but with 
associated latent vector $\hb^\prime\mbox{\boldmath $\eta$}$. 

\noindent{\it Theorem A.31.} If $\lambda_1$ and $\lambda_2$ are two distinct latent 
roots of a symmetric matrix $\hb$ of order $n$, then the corresponding latent 
vectors $\mbox{\boldmath $\eta$}_1$ and $\mbox{\boldmath $\eta$}_2$ are orthogonal.

\noindent{\it Proof.} Since $\hb$ is symmetric and 
$\hb\mbox{\boldmath $\eta$}_i=\lambda_i\mbox{\boldmath $\eta$}_i$ \ $(i=1,2)$, 
\begin{eqnarray*}
\lambda_2\mbox{\boldmath $\eta$}_1^\prime\mbox{\boldmath $\eta$}_2
&=\mbox{\boldmath $\eta$}_1^\prime(\lambda_2\mbox{\boldmath $\eta$}_2)&=\mbox{\boldmath $\eta$}^\prime_1\hb\mbox{\boldmath $\eta$}_2\\  
={}\mbox{\boldmath $\eta$}^\prime_2\hb\mbox{\boldmath $\eta$}_1
&=\mbox{\boldmath $\eta$}_2^\prime(\lambda_1\mbox{\boldmath $\eta$}_1)&=\lambda_1\mbox{\boldmath $\eta$}_2^\prime\mbox{\boldmath $\eta$}_1=\lambda_1
\mbox{\boldmath $\eta$}_1^\prime\mbox{\boldmath $\eta$}_2\;,
\end{eqnarray*}
\[{\rm i.e.} 
\qquad\lambda_1\mbox{\boldmath $\eta$}_1^\prime\mbox{\boldmath $\eta$}_2=\lambda_2\mbox{\boldmath $\eta$}_1^\prime\mbox{\boldmath $\eta$}_2\;.\]
Then since $\lambda_1\ne\lambda_2$, \ $\mbox{\boldmath $\eta$}_1^\prime\mbox{\boldmath $\eta$}_2=0$. 

\noindent{\it Theorem A.32.} If $\lambda$ is a latent root of a symmetric matrix $\hb$ 
and has multiplicity $m>1$, then it is possible to find  a set of $m$ 
orthonormal vectors which are latent vectors of $\hb$. 

The following is a result of the previous two theorems:

\noindent{\it Theorem A.33.} Let $\hb$ be a symmetric matrix of order $n$ with latent 
roots $\lambda_i$ \ ($i=1,\ldots,n$). Then $\hb$ can be written in the form 
\[\hb=\gb\pLambda\gb=\sum_{i=1}^n\lambda_i\mbox{\boldmath $\eta$}_i\mbox{\boldmath $\eta$}^\prime_i\;,\]
where $\pLambda$ is the matrix $\diag(\lambda_1,\ldots,\lambda_n)$ and $\gb$ is 
an orthogonal matrix whose $i$th column $\mbox{\boldmath $\eta$}_i$ is the normalized latent 
vector associated with $\lambda_i$.

The above decomposition of $\hb$ into a linear combination of orthogonal 
idempotent matrices $\mbox{\boldmath $\eta$}_i\mbox{\boldmath $\eta$}_i^\prime$ \ ($i=1,\ldots,n$) is called the 
spectral decomposition of $\hb$. 

\noindent{\it Theorem A.34.} Let $\hb$ be a symmetric matrix of order $n$ with latent 
roots $\lambda_i$ \ ($i=1,\ldots,n$). Then 
\begin{eqnarray*}
{\rm (i)}\quad\trace(\hb)&=&\sum_{i=1}^n\lambda_i\;,\\
{\rm (ii)}\quad\det(\hb)&=&\prod_{i=1}^n\lambda_i\;.
\end{eqnarray*}

\noindent{\it Proof.} (i) From Theorem A.33, $\hb$ can be written in the 
spectral form $\hb=\gb\pLambda\gb^\prime$. Thus 
\begin{eqnarray*} \trace(\hb)
&=&\trace(\gb\pLambda\gb^\prime)\\
&=&\trace(\pLambda\gb^\prime\gb)\\
&=&\trace(\pLambda)\;,\quad\mbox{(since $\gb$ is orthogonal)}\\
&=&\sum_{i=1}^n\lambda_i\;.
\end{eqnarray*}

(ii) In a similar fashion 
\begin{eqnarray*} \det(\hb)
&=&\det(\gb\pLambda\gb^\prime)\\
&=&\det(\pLambda\gb^\prime\gb)\\
&=&\det(\pLambda)\det(\gb^\prime\gb)\\
&=&\det(\pLambda)\\
&=&\prod_{i=1}^n\lambda_i\;.
\end{eqnarray*}
 
\noindent{\it Theorem A.35.} A symmetric matrix is non-singular if and only if its 
latent roots are all non-zero. 

\noindent{\it Theorem A.36.} The rank of a symmetric matrix $\hb$ is equal to the 
number of non-zero latent roots of $\hb$.

\noindent{\it Proof.} Suppose $\hb$ has $p$ non-zero latent roots 
$\lambda_1,\ldots,\lambda_p$ with corresponding orthonormal latent vectors 
$\mbox{\boldmath $\eta$}_1,\ldots,\mbox{\boldmath $\eta$}_p$. From Theorem A.33
\[\hb=\sum_{i=1}^p\lambda_i\mbox{\boldmath $\eta$}_i\mbox{\boldmath $\eta$}_i^\prime\;,\]
and hence $\mbox{\boldmath $\eta$}_1,\ldots,\mbox{\boldmath $\eta$}_p$ form an orthonormal basis for $\sr(\hb)$. 
Thus from Theorem A.13, 
\[{\rm rank}\,(\hb)=p\;.\]

\noindent{\it Theorem A.37.} The latent roots of a symmetric idempotent matrix $\hb$ of 
order $n$ are either zero or one. Hence rank$(\hb)=\trace(\hb)$. 

\noindent{\it Proof.} From Theorem A.33, $\hb$ can be written in the spectral 
form 
\[\hb=\gb\pLambda\gb^\prime\;,\]
and so $\hb^2=\gb\pLambda\gb^\prime\gb\pLambda\gb=\gb\pLambda^2\gb^\prime$, 
thus $\gb\pLambda\gb^\prime=\gb\pLambda^2\gb^\prime$.
Pre- and post-multiplication by $\gb^\prime$ and $\gb$ respectively gives 
$\pLambda=\pLambda^2$. In other words 
\[\lambda_i=\lambda_i^2\qquad(i=1,\ldots,n)\;.\]
Hence the $\lambda_i$ are either 0 or 1.

\noindent{\it Theorem A.38.} Suppose a symmetric matrix $\hb$ of order $n$ with latent 
roots $\lambda_1,\ldots,\lambda_n$ is written in the spectral form 
$\hb=\gb\pLambda\gb^\prime$, where $\pLambda=\diag(\lambda_1,\ldots,\lambda_n)$ 
and $\gb$ is an orthogonal matrix. A generalized inverse of $\hb$ is given by 
$\gb{\bf \Theta}\gb^\prime$, where ${\bf \Theta}$ is a diagonal matrix with elements 
${1\over \lambda_i}$ if $\lambda_i\ne 0$ and zero if $\lambda_i=0$ \ 
($i=1,\ldots,n$).

Such a generalized inverse of $\hb$ is called a Moore-Penrose generalized 
inverse and is denoted by $\hb^+$. If none of the latent roots of $\hb$ are 
zero, i.e.\ $\hb$ is non-singular then $\hb^+$ becomes the ordinary inverse, 
$\hb^{-1}$ of $\hb$.

\end{document}